\documentclass[useAMS,usenatbib]{mn2e}
\usepackage{times}
\usepackage[dvips]{graphicx}
\usepackage{mathrsfs}

\title[Radio-loud AGN: accretion mode and environment]{The link between accretion mode and environment in radio-loud active galaxies}
\author[J. Ineson et al.]
{J.~Ineson,$^{1}$
J.~H.~Croston,$^{1,2}$
M.~J.~Hardcastle,$^{3}$
R.~P.~Kraft,$^{4}$
D.~A.~Evans,$^{4}$
\newauthor
and M.~Jarvis$^{5,3,6}$ \\
  $^1$School of Physics and Astronomy, University of Southampton, Southampton SO17 1BJ, UK\\
  $^2$Institute of Continuing Education, University of Cambridge, Madingley Hall, Madingley, CB23 8AQ, UK.\\
  $^3$School of Physics, Astronomy and Mathematics, University of Hertfordshire, Hatfield, Hertfordshire AL10 9AB, UK \\
  $^4$Harvard-Smithsonian Center for Astrophysics, 60 Garden Street, Cambridge, MA 02138, USA \\
  $^5$Astrophysics, Department of Physics, Keble Road, Oxford OX1 3RH, UK\\
  $^6$Department of Physics, University of the Western Cape, Private Bag X17, Bellville 7535, South Africa }

\begin{document}

\date{Accepted 20xx xxxx xx. Received 2015 xxxx xx; in original form 2015 February xx}

\pagerange{\pageref{firstpage}--\pageref{lastpage}} \pubyear{20xx}

\maketitle

\label{firstpage}

\begin{abstract}
The interactions between radio-loud AGN and their environments play an important r\^{o}le in galaxy and cluster evolution. Recent work has demonstrated fundamental differences between High and Low Excitation Radio Galaxies (HERGs and LERGs), and shown that they may have different relationships with their environments. In the Chandra Large Project ERA (Environments of Radio-loud AGN), we made the first systematic X-ray environmental study of the cluster environments of radio galaxies at a single epoch ($z\sim0.5$), and found tentative evidence for a correlation between radio luminosity and cluster X-ray luminosity. We also found that this relationship appeared to be driven by the LERG sub-population \citep{ine13}.

We have now repeated the analysis with a low redshift sample ($z\sim0.1$), and found strong correlations between radio luminosity and environment richness and between radio luminosity and central density for the LERGs but not for the HERGs. These results are consistent with models in which the HERGs are fuelled from accretion discs maintained from local reservoirs of gas, while LERGs are fuelled more directly by gas ingested from the intra-cluster medium.

Comparing the samples, we found that although the maximum environment richness of the HERG environments is similar in both samples, there are poorer HERG environments in the $z\sim0.1$ sample than in the $z\sim0.5$ sample. We have therefore tentative evidence of evolution of the HERG environments. We found no differences between the LERG sub-samples for the two epochs, as would be expected if radio and cluster luminosity are related. 

\end{abstract}


\section{Introduction}\label{sec:intro}

The feedback cycles of radio-loud AGN play an important r\^{o}le in galaxy evolution (e.g. \citealt{cro06,sij06,mar14,scy15}), and we need to understand the processes that affect their behaviour in order to model galaxy formation and development. Over the past few years, compelling evidence has been uncovered for the importance of the dichotomy in the spectral types of radio-loud AGN --- the High and Low Excitation Radio Galaxies (HERGs, also known as radio-loud radiative mode AGN, and LERGs, or jet-mode AGN) --- and their different processes, environments and r\^{o}les in galaxy evolution. The current state of play is covered comprehensively in the reviews by \citet{mcn12} and \citet{hec14}, so we will restrict ourselves to a brief description of the observational evidence and processes of interest for this paper.

The two types of radio galaxy have different accretion rates (e.g. \citealt{bes12,son12,min14,gur14, fer15}), with the majority of HERGs having Eddington ratios of $\ga~1$~per~cent and LERGs of $\la~1$~per~cent, albeit with some overlap. The host galaxies display a different evolutionary phase, with HERGs tending to have lower mass than LERGs (e.g. \citealt{tas08, smo09,bes12}), to be bluer (e.g. \citealt{smo09,jan12,her10}), and to have ongoing star formation (eg. \citealt{har13}).

The dichotomy in accretion rate appears to be related to different accretion methods. HERGs, in common with radio-quiet radiative mode AGN, are thought to be fuelled by cold gas which builds some form of accretion disc and gives rise to radiatively efficient accretion \citep{sha73}. The radiation from the accretion disc indirectly gives rise to X-ray emission from the nucleus and IR from the surrounding dusty torus. Cold fuel is channelled into the galaxy centre, and because the host galaxies have a plentiful supply of cold gas, there is also fuel available for star formation. The high accretion rate adds mass to the black hole, and radiative mode AGN are thought to be responsible for black hole growth and the bulge-black~hole mass relation (e.g. \citealt{ish12}).

Galaxy mergers and interactions are thought to be a major source of new gas for HERGs. The predicted galaxy merger rates of \citet{hop08} reproduced the observed population properties of QSOs -- luminosity functions, fractions and  clustering -- suggesting that mergers and interactions could be a triggering and fueling mechanism. \citet{alm12} looked for signs of interactions in HERGs and comparable quiescent galaxies and found a substantially higher proportion of HERGs than quiescent galaxies with  comparably bright interaction features (90~per~cent vs 50~per~cent), implying that interactions are likely to be part of the triggering process. (Since interaction features endure for substantially longer than the active phase of a HERG, they calculated that we should expect HERGs in 1-10~per~cent of disturbed ellipticals.) \citet{tad14} estimated that gas reservoirs of around $10^9 M_{\odot}$ are required to sustain activity over the expected duty cycle of a HERG. They found that this matched observed HERG gas reservoirs whereas the reservoirs for quiescent galaxies are, on average, lower. The fact that there are also quiescent galaxies with large gas reservoirs suggests that the size of the reservoir alone is not sufficient to trigger activity, and that gas distribution and kinematics also play a part.

The fuelling mechanism for LERGs is more uncertain. They do not show the characteristic optical and X-ray features of radiatively efficient accretion seen in radiative mode AGN, and it is likely that they are fuelled by a hot, radiatively inefficient flow or advection dominated accretion flow (ADAF, e.g. \citealt{nar94}). Bondi accretion of hot gas could provide enough energy for the lower luminosity population \citep{all06,har07b}, so such sources could be fuelled by a form of hot flow, but this is probably insufficient for the more luminous sources (\citealt{rus13}, -- although this is disputed by \citealt{fuj14}). A number of researchers (e.g. \citealt{piz05,gas12,gas13,voi15}) discuss a process of chaotic accretion of cool gas clouds, which they find to be capable of releasing more than sufficient energy for the most powerful sources. In this model, filaments and blobs of cooling gas are channelled in from the ICM at large radii to form a central, rotating torus, and these gas clumps then collide and accrete, with both processes releasing energy. The cycle is controlled by the central entropy, which reduces the gas movement into the centre when high, but then drops again as the cold gas in the centre is accreted and removed by the jets. This allows the gas inflow to increase. Temperatures, accretion rates and jet powers thus vary (and in the case of lower power systems may turn off) during a well-regulated cycle that can be maintained for perhaps $10^8$ years.

Looking now at the cluster environments of the radio-loud AGN, the radio jets transport energy a considerable distance into the cluster, and are themselves affected by the intra-cluster medium. A common feature of all proposed LERG fuelling models is that they obtain material from within the ICM, so do the properties of the large-scale cluster environment in their turn affect the feedback loop maintaining the AGN, or are the AGN properties solely determined by the more local environment of the host galaxy? And how does does this disruption of the cluster environment affect its evolution?

It has been suggested for some time that there is a difference between the environments of HERGs and LERGs. At lower redshifts ($z \la 0.4$), the consensus of a number of studies is that HERGs appear to occupy a narrow range of relatively sparse environments compared with LERGs (e.g. \citealt{bes04,har04,gen13}). There is also some evidence that HERGs occupied richer environments in earlier epochs (e.g. \citealt{harv01,bel07,alm13}), suggesting evolution of the large-scale environment; conversely, \citet{wol00} and \citet{mcl01} found no evidence of evolution.

There are also mixed results when looking for relationships between radio luminosity and the cluster environment. \citet{bes04} found a correlation for LERGs but not for HERGs in the lower redshift range, as did \citet{ine13} at $z\sim0.5$. \citet{bel07} found no correlation for a mixed sample of mainly HERGs at higher redshifts, while \citet{wol00}, again at higher redshifts, found a correlation for QSOs.

There are a number of different measures of cluster richness, so comparing results needs some caution. \citet{bel07} and \citet{ine13} used X-ray luminosity within the virial radius and $R_{500}$ respectively. The other studies cited above all used optical methods -- a variation on the `nearest neighbours' method \citep{bes04}; galaxy counts above a specified magnitude within 1~Mpc radius \citep{gen13}; and the galaxy-galaxy spatial covariance function $B_{gg}$ calculated within a variety of radii depending on the image size (\citealt{harv01,har04,wol00} use 0.5~Mpc; \citealt{mcl01} use 180 kpc; \citealt{alm13} use 170 kpc). The different measures of cluster richness correlate, (e.g. \citealt{wol00,yee03,led03,ine13}), but with a large amount of scatter so conversion between the methods is unreliable. 

The different sizes of galaxy count regions may also present a problem. \citet{tas08} calculated galaxy overdensities within three different radii (75 kpc, 250 kpc and 450 kpc). Differences between 250 and 450 kpc were present but not strong, but the results for 75 kpc, which would capture interacting galaxies, were strikingly different from the larger-scale radii. The HERG-dominated population in particular was very overdense within 75 kpc but underdense within the higher radii. Studies using galaxy count regions with large and small radii may therefore be capturing different effects. 

Furthermore, if there is evolution of the environment, studies of environment richness at high and low redshift may not be directly comparable and studies covering a wide redshift range are at risk of confounding factors. In addition, the Malmquist bias in flux-limited samples, the changes in HERG and LERG populations with redshift and the paucity of powerful local sources also need to be taken into account during sample selection and data analysis. The studies cited above all agree on the differences between HERG and LERG environments at relatively low redshifts; the disagreements come when making comparisons across wider redshift ranges.

In the ERA (Environments of Radio-loud AGN) programme, we are making a systematic examination of the effects of epoch and environment on the properties of radio-loud AGN in order to address the two questions separately: is radio luminosity is related to the large-scale cluster environment? and does the environment evolve with epoch? In phase 1 (\citealt{ine13}, I13 hereafter), we used the ERA sample (Figure~\ref{fig:zLr}) to compare radio luminosity and the hot gas environment within a limited redshift range $(0.4<z<0.6)$, thus removing the effects of redshift evolution. We found that the LERGs occupied a wide range of environments, and that there was a correlation between radio luminosity and environment richness. In contrast, the HERGs occupied a smaller range of environment richnesses and showed no sign of a correlation. \citet{her15}, using the galaxy-quasar covariance function rather than the X-ray luminosity of the ICM found a similar result in the same redshift range, as did \citet{bes04} with a lower redshift sample ($z<0.1$).

In this paper, we continue the ERA programme by making a similar analysis of a low redshift comparison sample with $z\sim0.1$, to see if the results from the ERA sample carry across to the lower redshift, and if there is any evidence of environment change with epoch.

Throughout this paper we use a cosmology in which $H_0 = 70$ km s$^{-1}$ Mpc$^{-1}$, $\Omega_{\rm m} = 0.3$ and $\Omega_\Lambda = 0.7$. Unless otherwise stated, errors are quoted at the 1$\sigma$ level.

\section{The sample}

The new sample, hereafter referred to as the z0.1 sample, consists of all radio-loud AGN with visible structure beyond the nucleus lying in the redshift range $0.01\leq z\leq 0.2$ from two flux-limited radio surveys: the 3CRR survey \citep{lai83} and the sub-sample of the 2Jy survey (\citealt{wal85,tad93}) defined by \citet{dic08}. The initial sample contained 38 3CRR and 22 2Jy sources, but five sources were excluded. 3C~382 has currently no suitable {\it Chandra} or {\it XMM-Newton} observation, 3C~83.1B, 3C~264 and Abell~1552 lie in the outskirts of richer clusters, so that their immediate environments could not be disentangled from those of the stronger sources for the spherical modelling, while the angular size of the cluster emission of 3C~84 (Perseus~A) is so large that analysis was impractical. The final sample consists of 55 sources comprising 25 HERGs and 30 LERGS. 22 of the sources have FRI morphologies, and 33 are FRIIs.

\begin{figure*}
  \begin{minipage}{8cm}
  \includegraphics[width=7cm]{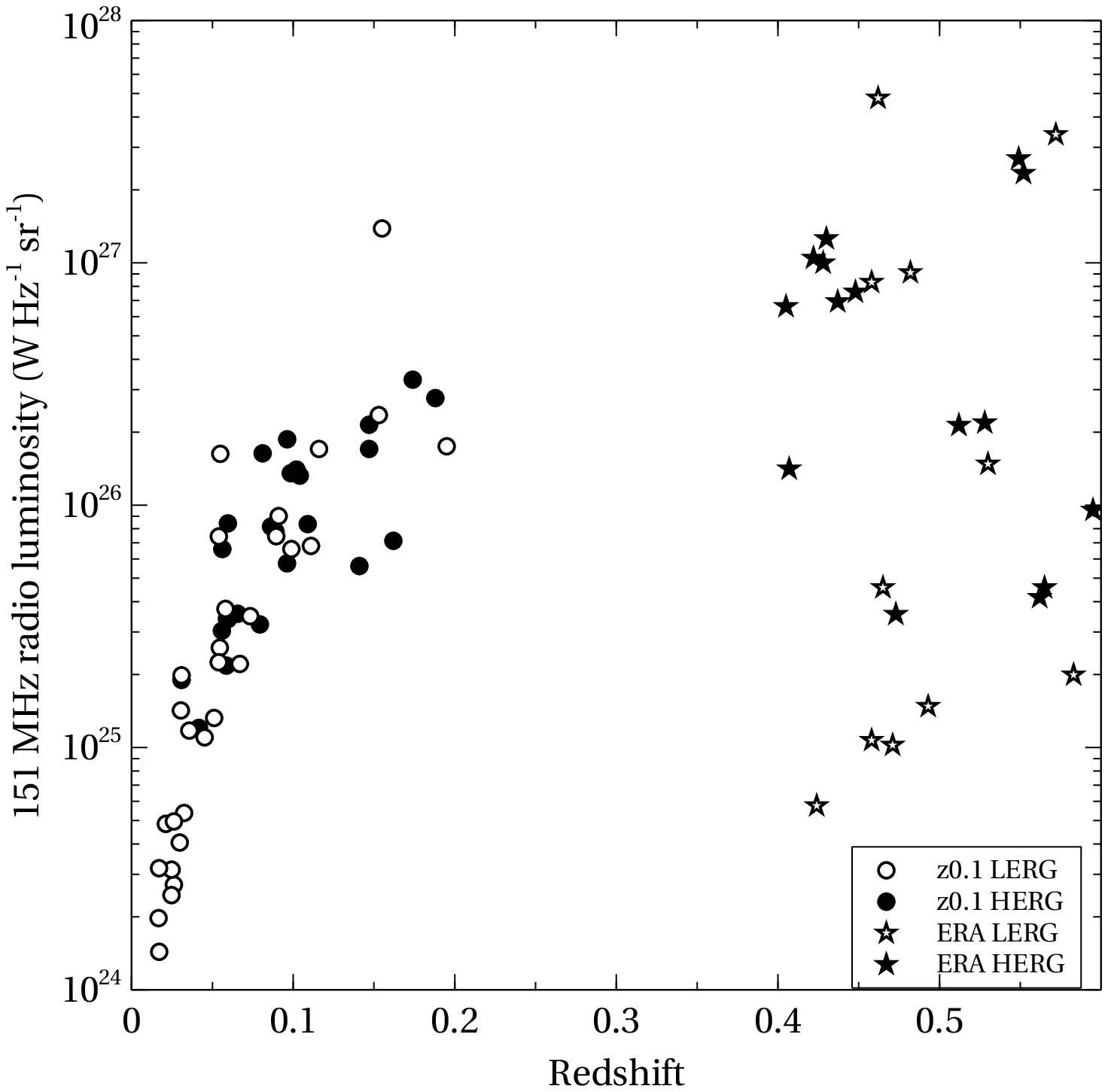}
  \end{minipage}
  \begin{minipage}{8cm}
  \includegraphics[width=7cm]{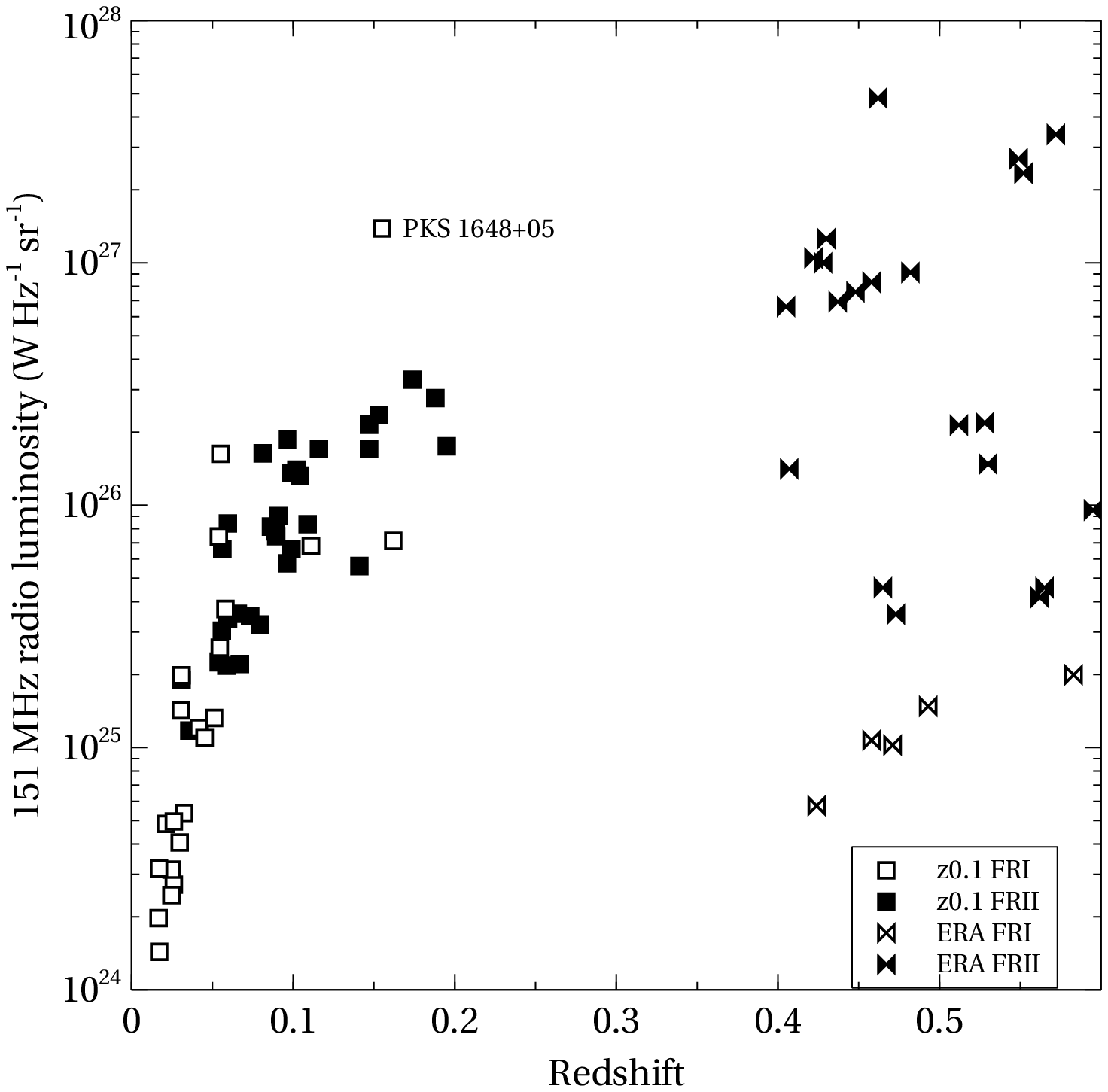}
  \end{minipage}
  \caption{151~MHz radio luminosity vs redshift for the z0.1 sample and the ERA ($z \sim 0.5$) sample used in I13. The left plot shows the HERG (filled symbols) and LERG (empty symbols) excitation classes and the right plot shows FRI (empty symbols) and FRII (filled symbols) morphologies. PKS~1648+05 has characteristics of both morphologies.}
\label{fig:zLr}\end{figure*}

The properties of the sources are listed in Table~\ref{tab:rawdata} and Figure~\ref{fig:zLr} shows radio luminosity plotted against redshift. The ERA sample sources used in I13 are also shown in Figure~\ref{fig:zLr}. Positions, radio luminosities, spectral type and morphologies of the sources were taken from the on-line 3CRR catalogue\footnote{http://3crr.extragalactic.info/} and from \citet{min14}. Redshifts were taken from the most recent source cited in the 3CRR catalogue, \citet{min14}, the NASA/IPAC Extragalactic Database (NED)\footnote{http://ned.ipac.caltech.edu/} and the SIMBAD astronomical database\footnote{http://simbad.u-strasbg.fr/simbad/}. Galactic column densities came from \citet{dic90} via the \textsc{heasarc} tools.

\begin{table*}
\caption{The z0.1 sample}
\begin{tabular}{lllcccccc}\hline

Source&\multicolumn{2}{c}{RA  (J2000)   Dec}&Redshift&Scale&log$_{10}$ L$_{151}$&Type&FR&nH\\
& h m s &deg min sec&&kpc arcsec$^{-1}$&W~Hz$^{-1}$~sr$^{-1}$&&Morphology&x10$^{20}$ cm$^{-2}$\\
\hline
3C 28&00 55 50.65&+26 24 37.3&0.195&3.23&26.24&LERG&2&5.39\\
3C 31&01 07 24.96&+32 24 45.2&0.017&0.34&24.30&LERG&1&5.36\\
3C 33&01 08 52.86&+13 20 14.2&0.060&1.15&25.93&HERG&2&3.90\\
3C 35&01 12 02.26&+49 28 35.5&0.067&1.28&25.34&LERG&2&13.00\\
3C 66B&02 23 11.41&+42 59 31.5&0.021&0.43&24.68&LERG&1&9.15\\
3C 76.1&03 03 14.99&+16 26 19.0&0.033&0.65&24.73&LERG&1&10.60\\
3C 98&03 58 54.43&+10 26 02.8&0.031&0.62&25.28&HERG&2&15.10\\
3C 192&08 05 35.01&+24 09 49.7&0.060&1.15&25.53&HERG&2&4.21\\
3C 219&09 21 08.63&+45 38 57.3&0.174&2.95&26.52&HERG&2&1.51\\
3C 236&10 06 01.76&+34 54 10.2&0.099&1.83&25.82&LERG&2&1.23\\
3C 285&13 21 17.86&+42 35 14.8&0.079&1.50&25.51&HERG&2&1.27\\
3C 293&13 52 17.80&+31 26 46.4&0.045&0.89&25.04&LERG&1&1.29\\
3C 296&14 16 52.98&+10 48 27.2&0.025&0.50&24.50&LERG&1&1.88\\
3C 303&14 43 02.76&+52 01 37.2&0.141&2.48&25.75&HERG&2&1.58\\
3C 305&14 49 21.64&+63 16 14.0&0.042&0.82&25.08&HERG&1&1.70\\
3C 310&15 04 57.11&+26 00 58.3&0.054&1.05&25.87&LERG&1&3.70\\
3C 321&15 31 43.46&+24 04 19.0&0.096&1.78&25.76&HERG&2&4.11\\
3C 326&15 52 09.10&+20 05 48.3&0.089&1.67&25.87&LERG&2&3.81\\
3C 338&16 28 38.29&+39 33 04.2&0.031&0.62&25.30&LERG&1&0.86\\
3C 346&16 43 48.60&+17 15 49.4&0.162&2.79&25.85&HERG&1&5.67\\
3C 386&18 38 26.22&+17 11 50.2&0.017&0.35&24.50&LERG&1&18.10\\
3C 388&18 44 02.35&+45 33 29.6&0.091&1.70&25.95&LERG&2&6.47\\
3C 390.3&18 42 08.93&+79 46 17.2&0.056&1.09&25.82&HERG&2&4.27\\
3C 433&21 23 44.56&+25 04 28.0&0.102&1.88&26.15&HERG&2&11.90\\
3C 442A&22 14 46.88&+13 50 27.2&0.026&0.53&24.70&LERG&1&5.07\\
3C 449&22 31 20.62&+39 21 30.1&0.017&0.35&24.16&LERG&1&12.00\\
3C 452&22 45 48.75&+39 41 15.9&0.081&1.53&26.21&HERG&2&11.30\\
3C 465&23 38 29.36&+27 01 53.4&0.030&0.61&25.15&LERG&1&5.01\\
4C 73.08&09 49 45.78&+73 14 23.1&0.059&1.14&25.34&HERG&2&2.29\\
DA 240&07 48 36.82&+55 48 59.5&0.036&0.71&25.07&LERG&2&4.52\\
NGC 6109&16 17 40.56&+35 00 15.7&0.030&0.59&24.61&LERG&1&1.33\\
NGC 6251&16 32 31.95&+82 32 16.4&0.025&0.50&24.39&LERG&1&5.65\\
NGC 7385&22 49 54.59&+11 36 32.5&0.026&0.53&24.43&LERG&1&5.11\\
PKS 0034-01&00 37 49.18&-01 09 08.2&0.073&1.40&25.54&LERG&2&3.07\\
PKS 0038+09&00 40 50.53&+10 03 26.8&0.188&3.14&26.44&HERG&2&5.51\\
PKS 0043-42&00 46 17.75&-42 07 51.4&0.116&2.10&26.23&LERG&2&2.21\\
PKS 0213-13&02 15 37.5&-12 59 30.5&0.147&2.57&26.23&HERG&2&1.92\\
PKS 0349-27&03 51 35.81&-27 44 33.8&0.066&1.26&25.55&HERG&2&0.99\\
PKS 0404+03&04 07 16.49&+03 42 25.8&0.089&1.66&25.89&HERG&2&11.90\\
PKS 0442-28&04 44 37.67&-28 09 54.6&0.147&2.57&26.33&HERG&2&2.43\\
PKS 0620-52&06 21 43.29&-52 41 33.3&0.051&1.00&25.12&LERG&1&5.17\\
PKS 0625-35&06 27 6.65&-35 29 16.3&0.055&1.06&25.41&LERG&1&7.24\\
PKS 0625-53&06 26 20.44&-53 41 35.2&0.054&1.05&25.35&LERG&2&5.42\\
PKS 0806-10&08 08 53.600&-10 27 39.71&0.109&1.99&25.92&HERG&2&7.74\\
PKS 0915-11&09 18 05.67&-12 05 44.0&0.055&1.07&26.21&LERG&1&4.93\\
PKS 0945+07&09 47 45.15&+07 25 20.4&0.086&1.62&25.91&HERG&2&3.00\\
PKS 1559+02&16 02 27.38&+01 57 55.7&0.104&1.91&26.12&HERG&2&6.44\\
PKS 1648+05&16 51 08.16&+04 59 33.8&0.155&2.69&27.14&LERG&1&6.33\\
PKS 1733-56&17 37 35.80&-56 34 03.4&0.099&1.82&26.13&HERG&2&8.29\\
PKS 1839-48&18 43 14.64&-48 36 23.3&0.111&2.02&25.83&LERG&1&5.62\\
PKS 1949+02&19 52 15.77&+02 30 23.1&0.059&1.14&25.53&HERG&2&14.30\\
PKS 1954-55&19 58 16.64&-55 09 39.7&0.058&1.13&25.57&LERG&1&4.52\\
PKS 2211-17&22 14 25.78&-17 01 36.3&0.153&2.66&26.37&LERG&2&2.61\\
PKS 2221-02&22 23 49.57&-02 08 12.4&0.056&1.09&25.48&HERG&2&4.87\\
PKS 2356-61&23 59 04.50&-60 54 59.1&0.096&1.78&26.27&HERG&2&2.38\\
\hline\end{tabular}

\label{tab:rawdata}\end{table*}

\section{OBSERVATIONS AND DATA PREPARATION}

\subsection{X-ray data}\label{sec:Xray}

All but two of the X-ray observations for the z0.1 sample came from the {\it Chandra} archive; for 3C~31 and 3C~66B we used observations from the {\it XMM-Newton} archive. The {\it XMM-Newton} observations used the three EPIC cameras with the medium filter and the {\it Chandra} observations used either the ACIS-S3 chip or the four ACIS-I chips. Observation IDs and times are given in Table~\ref{tab:obsdata}.

Initial data processing for the {\it Chandra} observations was done in the same way as in I13, using the \textit{Chandra} analysis package \textsc{ciao} v4.5 and \textsc{caldb} v4.5.8. The files were reprocessed using \textit{chandra\_repro}, with particle background cleaning for observations in the VFAINT mode, and background flares were removed using \textit{deflare}. Screened observation times are included in Table~\ref{tab:obsdata}.

Since the clusters and groups at $z\sim0.1$ had a larger angular size than the more distant observations in I13, in some cases extending beyond the observing chip, two additional processes were used. In order to correct for vignetting, aspect, chip edges and gaps, we created exposure maps using the \textit{fluximage} script. We also built background files from the {\it Chandra} blank sky datasets for each observation using the method described in the \textsc{ciao} threads\footnote{http://cxc.harvard.edu/ciao/threads/acisbackground/}. These were used when generating surface brightness profiles and spectra. 

The \textit{XMM-Newton} events files were processed in the same way as in I13, using using \textit{XMM-Newton} \textsc{sas} v11.0.0. The reprocessed events files were checked for flares and then corrected for vignetting using \textit{evigweight}. Again, the particle background in the \textit{XMM-Newton} sources was removed using the method described by \citet{cro08a} using the closed filter files supplied courtesy of E. Pointecouteau.

Both the {\it Chandra} blank sky files and the \textit{XMM-Newton} closed filter files were scaled by comparing the source and background count rates at 10-12 keV (ACIS and MOS cameras) and 12-14 keV (pn camera). The background file count rates were scaled by this factor. Note that the {\it Chandra} blank sky files are cleaned, open filter datasets and so can be used to model the ACIS background as well as to remove instrumental background whereas the \textit{XMM-Newton} closed filter files remove only the instrumental background.

We checked for pileup by finding the maximum count rate per pixel at the peak of the emission, which we scaled to estimate the pileup fraction\footnote{http://cxc.harvard.edu/csc/memos/files/Davis\_pileup.pdf}. If pileup was greater than 12~per~cent, we checked the maximum count rate per pixel at increasing radii until pileup was less than 12~per~cent. We then excluded the central region out to this radius. One source had severe pileup (3C~390.3), and three had mild pileup (3C~219, 3C~303 and NGC~6251).

For both the \textit{Chandra} and \textit{XMM-Newton} sources, we examined images of the sources in \textit{ds9}, overlaying them with the radio contours. We identified point sources in the data sets and emission associated with the radio lobes, which we excluded during subsequent analysis.

\begin{table*}
\caption{Observation data for the z0.1 sample}
\begin{tabular}{lccccccl}\hline
Source&X-ray${^a}$&Observation&Exposure${^b}$&Screened${^b}$&Radio map&Resolution&Ref.\\
Instrument&&ID&time (ks)&time (ks)&freq. (GHz)&(arcsec)\\
\hline
3C 28&C&3233&50.38&49.30&1.4&$1.1 \times 1.1$&1\\
3C 31&XMM&551720101&50.00&24.00&0.61&$52 \times 29$&1\\
3C 33&C&6910, 7200&39.83&39.61&1.5&$4 \times 4$&1\\
3C 35&C&10240&25.63&25.63&1.4&$14 \times 12$&1\\
3C 66B&XMM&0002970201&17.86&13.39&1.4&$12 \times 12$&1\\
3C 76.1&C&9298&8.06&8.06&1.5&$4.9 \times 4.9$&1\\
3C 98&C&10234&31.71&31.71&4.9&$3.7 \times 3.7$&1\\
3C 192&C&9270&10.02&9.62&1.4&$3.9 \times 3.9$&1\\
3C 219&C&827&19.24&16.79&1.5&$1.4 \times 1.4$&1\\
3C 236&C&10249&40.50&40.50&0.61&$28 \times 28$&1\\
3C 285&C&6911&39.62&39.61&1.5&$5.5 \times 5.5$&1\\
3C 293&C&12712&67.81&67.22&1.5&$7.6 \times 7.6$&1\\
3C 296&C&3968&49.43&48.91&1.5&$4.9 \times 4.9$&1\\
3C 303&C&1623&15.10&14.95&1.5&$1.2 \times 1.2$&1\\
3C 305&C&12797, 13211&57.32&57.31&1.5&$0.15 \times 0.15$&1\\
3C 310&C&11845&57.58&57.16&1.5&$4 \times 4$&1\\
3C 321&C&3138&47.13&46.87&1.5&$1.4 \times 1.4$&1\\
3C 326&C&10908, 10242&45.81&45.81&1.4&$14 \times 39$&1\\
3C 338&C&10748&40.58&40.58&4.9&$1 \times 1$&1\\
3C 346&C&3129&46.69&39.91&1.5&$0.35 \times 0.35$&1\\
3C 386&C&10232&29.29&29.29&1.5&$5.8 \times 5.8$&1\\
3C 388&C&5295&30.71&30.31&1.4&$1.3 \times 1.3$&1\\
3C 390.3&C&830&33.97&28.60&1.6&$2.8 \times 2.8$&1\\
3C 433&C&7881&38.19&37.15&8.5&$0.75 \times 0.75$&1\\
3C 442A&C&5635, 6353&40.99&40.79&1.4&$7.5 \times 7.5$&1\\
3C 449&C&4057&29.18&24.99&0.61&$30 \times 48$&1\\
3C 452&C&2195&79.92&79.53&1.4&$6 \times 6$&1\\
3C 465&C&4816&49.53&49.49&1.4&$5.4 \times 5.4$&1\\
4C 73.08&C&10239&28.52&28.52&0.61&$30 \times 30$&1\\
DA 240&C&10237&24.08&24.08&0.61&$34 \times 34$&1\\
NGC 6109&C&3985&19.39&13.42&1.4&$13 \times 13$&1\\
NGC 6251&C&847&37.44&30.67&0.33&$55 \times 55$&1\\
NGC 7385&C&10233&39.33&39.33&4.9&$4.9 \times 3.6$&2\\
PKS 0034-01&C&2178&27.52&26.54&4.9&$4.5 \times 3.7$&3\\
PKS 0038+09&C&9293&7.94&7.94&4.9&$4.4 \times 3.4$&3\\
PKS 0043-42&C&10319&18.38&18.38&8.6&$1.2 \times 0.88$&4\\
PKS 0213-13&C&10320&19.89&19.89&4.9&$5.9 \times 3.4$&3\\
PKS 0349-27&C&11497&19.89&19.89&1.5&$11 \times 8.9$&2\\
PKS 0404+03&C&9299&8.07&8.07&8.4&$2.2 \times 2.2$&5\\
PKS 0442-28&C&11498&19.79&19.79&4.9&$9.3 \times 2.1$&3\\
PKS 0620-52&C&11499&19.80&19.80&4.7&$2.6 \times 1.5$&3\\
PKS 0625-35&C&11500&19.79&19.79&4.9&$4.7 \times 3.2$&2\\
PKS 0625-53&C&4943&18.45&16.67&4.8&$2 \times 1.6$&4\\
PKS 0806-10&C&11501&19.79&19.79&4.9&$6.8 \times 1.6$&2\\
PKS 0915-11&C&4970&98.82&98.42&1.4&$2 \times 1.5$&2\\
PKS 0945+07&C&6842&29.78&29.78&1.5&$4 \times 4$&6\\
PKS 1559+02&C&6841&39.65&39.62&8.5&$2.2 \times 2.2$&5\\
PKS 1648+05&C&6257&49.52&49.52&1.5&$1.4 \times 1.4$&7\\
PKS 1733-56&C&11502&19.86&19.66&4.7&$2.2 \times 1.3$&3\\
PKS 1839-48&C&10321&19.79&19.79&4.7&$2.6 \times 1.7$&3\\
PKS 1949+02&C&2968&49.47&45.66&1.5&$4.5 \times 4.1$&8\\
PKS 1954-55&C&11505&20.65&20.45&4.8&$2.4 \times 1.3$&4\\
PKS 2211-17&C&15091&164.38&163.17&4.9&$7.6 \times 3.1$&3\\
PKS 2221-02&C&7869&45.60&45.60&8.2&$2.4 \times 2.4$&5\\
PKS 2356-61&C&11507&19.79&19.79&1.5&$7.2 \times 6.9$&9\\
\hline\end{tabular}

References: (1) http://www.jb.man.ac.uk/atlas, (2) Made from the VLA archives, (3) Morganti et al. (1993), (4) Morganti et al. (1999), (5) Leahy et al. (1997), (6) Hardcastle et al. (2007a), (7) Gizani \& Leahy (2003), (8) Dennett-Thorpe et al. (2002), (9) Made from the ATCA archives.

${^a}$ C=\textit{Chandra}, X=\textit{XMM-Newton}.
${^b}$ pn camera times for {\it XMM-Newton} sources.
\label{tab:obsdata}\end{table*}

\subsection{Radio data}

Radio maps were used to mask out the radio lobes so that any radio-related X-ray emission did not contaminate our measurements of the cluster properties. All of the 3CRR radio maps except that of NGC~7385 were taken from the 3CRR Atlas\footnote{http://www.jb.man.ac.uk/atlas}. We also used existing maps for the majority of the 2Jy sources. The remaining six maps were made using observations from the VLA and ATCA archives and reduced using {\sc aips} in the standard manner.

Table~\ref{tab:obsdata} contains the full details of the radio maps used, including references.

\section{ANALYSIS}

The aim of the analysis was to find the temperature and X-ray luminosity of the ICM emission of the radio galaxies. Where possible, the temperature was obtained by spectral analysis; when there were insufficient counts, it was estimated from the count rate in a self-consistent way, as described in Section~\ref{sec:spat}. The luminosity was determined by integrating the surface brightness profiles to the $R_{500}$ overdensity radius (defined in Section~\ref{sec:spat}).

The appendix contains brief notes on the individual sources.

\subsection{Spatial analysis}\label{sec:spat}

We extracted a radial surface brightness profile from the events file of each source by taking the average counts in annuli around the source centroid. The point sources and radio emission identified during data preparation were removed and the annulus areas adjusted to account for the excluded regions. We used an energy range of 0.4-7.0 keV, this being the well calibrated range for the \textit{Chandra} data. When possible, we used a double subtraction method where the background in the blank sky/closed filter files was subtracted from the source file (see \citealt{mar03}) prior to generating the profile. Then, when the maximum detection radius lay at least partially within the ACIS-S3 or ACIS-I chips, we used an annulus outside the maximum detection radius to obtain the background counts remaining after the initial background subtraction, and these were subtracted from the source counts to obtain the net counts in each annulus. 

For the \textit{XMM-Newton} sources, since the pn camera is more sensitive than the MOS cameras, we created the pn profile first and then used the same annuli and background area for the MOS profiles. The three profiles were then scaled by their relative exposure times and added together.

Table~\ref{tab:beta} contains the maximum detection radius and net counts within that radius for each of the sources. If the emission extends beyond the chip, the maximum detection radius is quoted as the distance to the chip edge.

We fitted the surface brightness profiles with $\beta$ models \citep{cav76} in a similar manner to that described in I13, using the Markov-Chain Monte Carlo (MCMC) method described by \citet{cro08a} to explore the parameter space of these models and find Bayesian estimates of the core radii ($r_{c}$) and $\beta$ values. However, since these sources are nearer than those in I13, their host galaxies usually extend beyond the instrument point spread function (PSF). In these cases, a second $\beta$ model was added so that the inner and outer components were modelled individually.

We initially checked whether the PSF alone gave a satisfactory fit to the data, which was not the case for any of the sources. We then added a single $\beta$ model to fit the extended emission, convolved with the PSF. The unconvolved surface brightness at radius $R$ is given by
\begin{equation}
S(R)=S_0(1+(\frac{R}{r_c})^2)^{-3\beta+0.5}
\end{equation}
\noindent where $r_c$ is the core radius and $S_0$ is the normalisation. 

If this gave a satisfactory fit, we checked whether the modelled emission extended beyond the host galaxy radius (taken from NED\footnote{http://ned.ipac.caltech.edu/}). If it did not, which was the case for seven sources, we used the observations of the source to derive an upper limit on cluster-scale emission. For these sources, we compared the net count rate from a region outside the host galaxy with a more distant background region. If the net counts were greater than three times the background error, we considered this to be a detection; otherwise we used three times the background error as the upper limit. The latter was the case for all seven cases. 

If the single $\beta$ model did not give a satisfactory fit we added the second $\beta$ model as described in \citet{cro08a}, so that we could describe the galaxy and cluster contributions separately and exclude the galaxy luminosity. In this model, the line-of-sight sum of the gas densities in the two components is given by
\begin{equation}
n(r)=n_0\Big[\Big(1+(\frac{r}{r_{c,in}})^2\Big)^{-3\beta_{in}/2}+N\Big(1+(\frac{r}{r_c})^2\Big)^{-3\beta/2}\Big]
\end{equation}
\noindent where $N$ is the relative normalisation of the two beta models and the subscript $in$ refers to the inner (galaxy) model.

The surface brightness profile is then
\begin{equation}
S(R)\propto\int_{-\infty}^\infty n^2(l,R)dl
\end{equation}
\noindent where the radius $r=\sqrt{l^2+R^2}$, and $l$ is the distance along the line of sight.

Since we are assuming the clusters and groups to be at least approximately relaxed, we limited $\beta$ for the outer (ICM) model to the range 0.3 to 1.2 and $r_c$ to a minimum of 1~kpc. We allowed more freedom in the parameters for the inner model, which describes the host galaxy, since the galaxy is ellipsoidal rather than spherical and could be at any orientation to the line of sight.
   
\begin{figure}
  \begin{minipage}{8cm}
  \includegraphics[width=7cm]{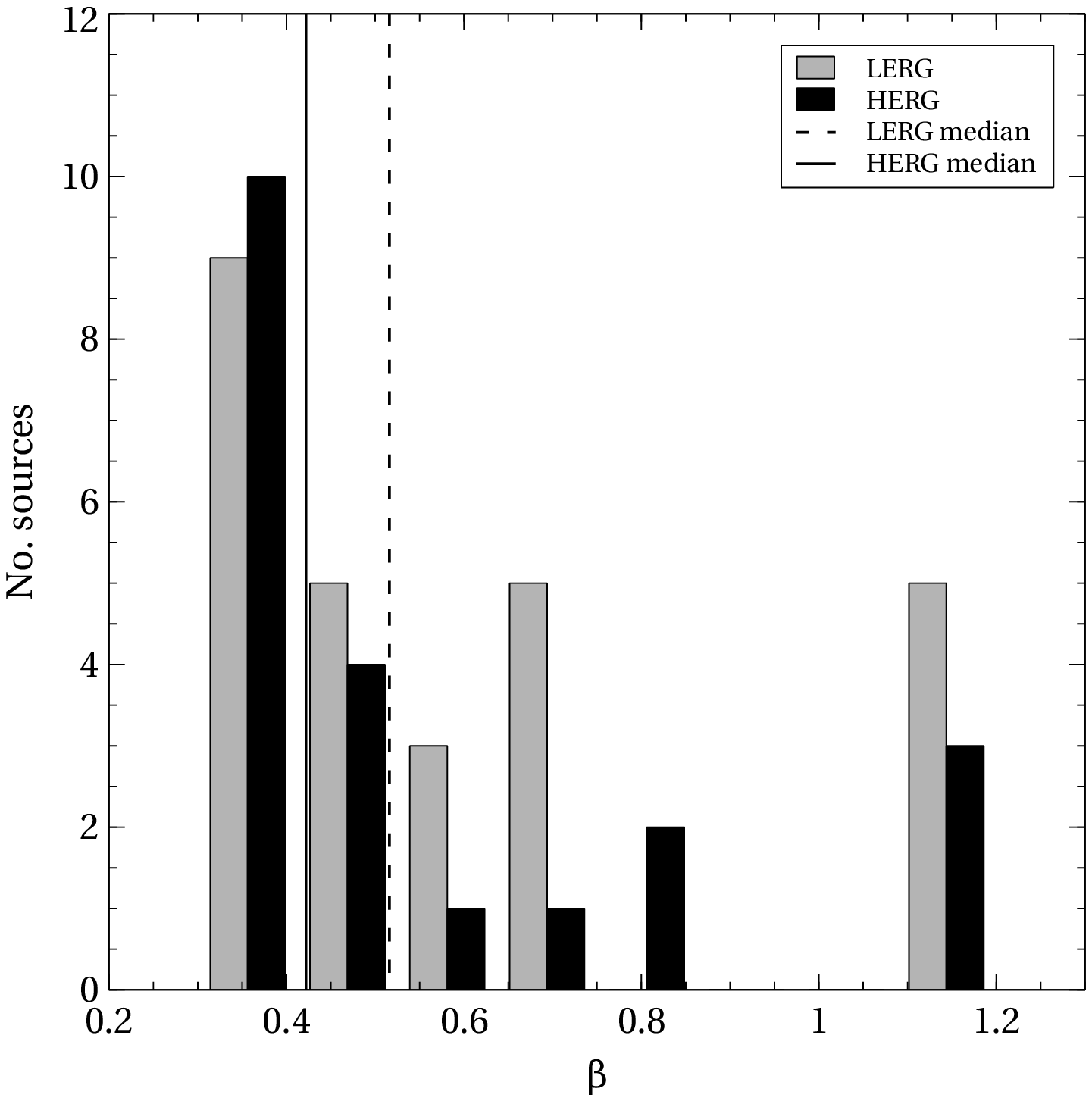}
  \end{minipage}
  \begin{minipage}{8cm}
  \includegraphics[width=7cm]{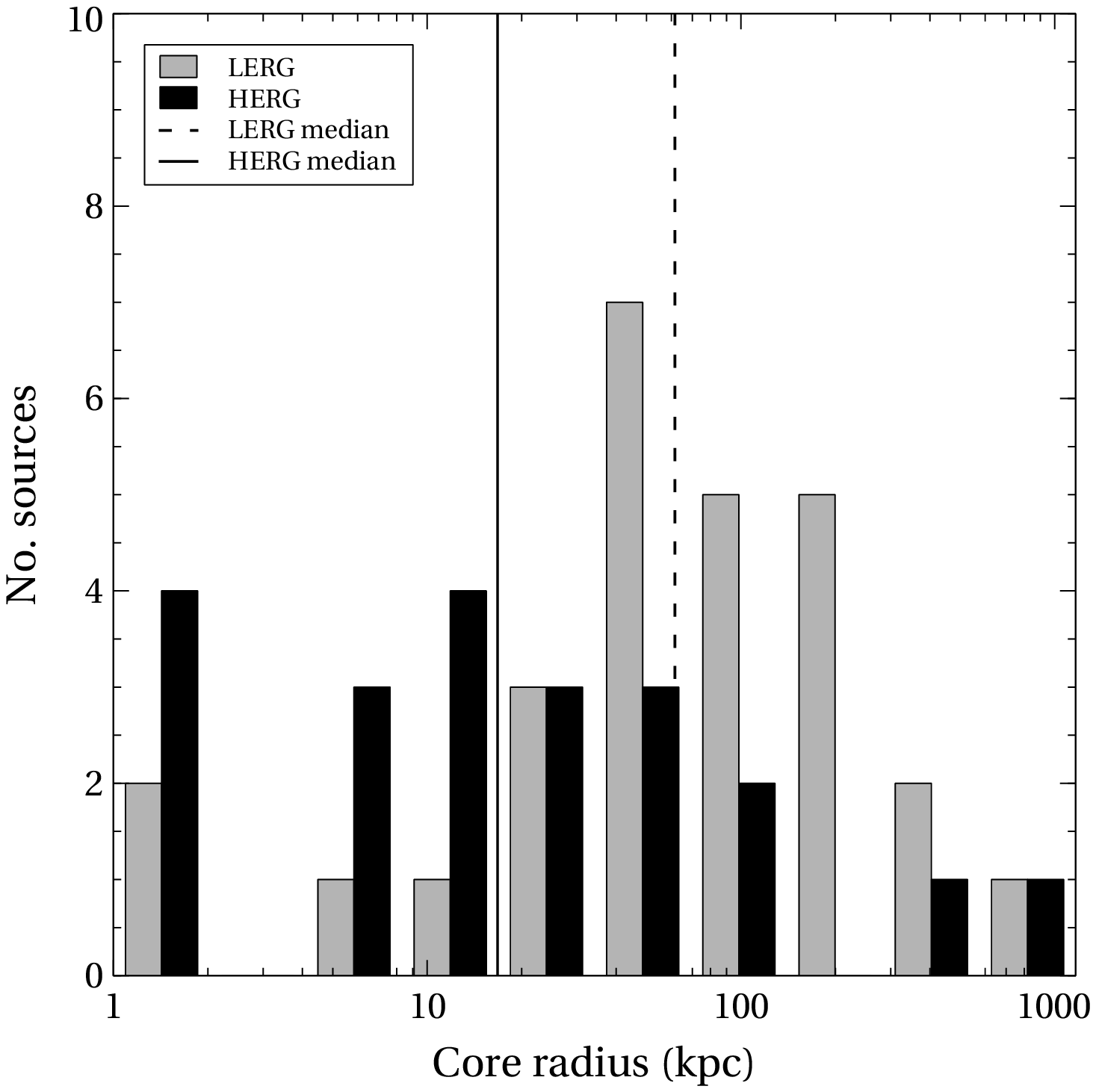}
  \end{minipage}
  \caption{Distribution of the $\beta$ model parameters for the z0.1 sample; $\beta$ (above) and core radius (below). The LERGs are grey and the HERGs black. Dashed lines show the LERG medians and solid lines the HERG medians.}
\label{fig:betas}\end{figure}

The goodness-of-fit and ICM $\beta$ model parameters are shown in Table~\ref{tab:beta}, and Table~\ref{tab:betain} gives the inner beta model parameters for the sources where the galaxy emission was discernible. The distributions of $\beta$ and $r_c$ for the ICM model are shown in Figure~\ref{fig:betas}. Except for a few high values, the values of $\beta$ are mostly lower than the values of typical clusters, as is expected for less rich environments (e.g. \citealt{mul00}). The overall median $\beta$ for the z0.1 sample is 0.47, close to the value of 0.5 expected for groups. The ERA sample also has a median of 0.47, showing consistency between the two samples.

There are eight sources with high values of $\beta$ compared with the rest of the sample and six sources with a very low core radius. 3C~442A and PKS~0915$-$11 (Hyd~A) contain interacting galaxies resulting in shocks propagating through the ICM. The disturbance is clearly visible in the surface brightness profiles. Both profiles contain sufficient counts to show that the outer ICM is well modelled. The other sources have very little emission detected beyond the host galaxy and so their models are poorly constrained. The profiles of the environments with high $\beta$s have emission visible beyond the host, but because there are few bins the profiles look flat and wide, and this is reflected in the model parameters. The outer emission of those sources with low core radii have the opposite problem, with the outer profile merging smoothly with that of the host galaxy so they are hard to differentiate.

For some of the inner $\beta$ models, it was difficult to differentiate between the point source and the galaxy, and consequently some of the galaxy $\beta$s are very high. However, as we only wished to model the shapes in order to exclude them from the luminosity calculation, we did not expect this to be of concern provided the profile shape was well modelled. We tested this assumption using the sources with the two highest galaxy $\beta$s --- 3C~98 and PKS~1559+02. We fixed the galaxy $\beta$s at the median value of 0.95, refitted the surface brightness profiles and recalculated the luminosities. They were both consistent with the luminosities calculated using the high values of $\beta$.

Because the \textit{Chandra} blank sky files may not model the background accurately, we looked for systematic differences between the profiles made using double and single background subtraction. All the single subtraction sources have a large number of counts and well-defined profiles, and the $\beta$ model parameters cover a similar range to those of the double subtraction sources. All except PKS~1648+05 (Her~A) are towards the low end of the redshift range, so would be expected to cover the chip if the environment was relatively rich and/or the observation long.

\begin{table*}
\caption{Radial profile modeling - ICM (outer) $\beta$ model}
\begin{tabular}{llcccll}\hline
Source&Model$^{a}$,&$R_{det}$&Counts$^{c,d}$&$\chi^2$/dof&
  $\beta^{e,f}$&$r_c$\\
&Method$^{b}$&kpc&&&&kpc\\
\hline
3C 28&$d\beta$, dsub&1432&14810&74/15&0.67 (0.55--0.85)&340.12 (253.47--437.35)\\
3C 31&$d\beta$, dsub&203&10827&104/62&0.30 (0.30--1.50)&39.34 (27.73--256.55)\\
3C 33&$d\beta$, dsub&113&2770&3.2/8&0.76 (0.30--1.20)&16.76 (1.46--112.81)\\
3C 35&$s\beta$, dsub&221&218&3.5/5&1.17 (0.30--1.20)&134.53 (2.08--384.43)\\
3C 66B&$d\beta$, dsub&282&13788&13/14&0.35 (0.31--0.37)&46.91 (33.33--59.22)\\
3C 76.1&$s\beta$, dsub&95&117&0.6/2&0.46 (0.30--1.20)&9.84 (0.82--64.56)\\
3C 98&$d\beta$, dsub&121&1380&1.4/8&0.42 (0.30--1.20)&1.79 (1.00--58.13)\\
3C 192&$s\beta$, dsub&170&191&1.1/3&0.41 (0.30--0.91)&1.02 (1.00--10.63)\\
3C 219&$s\beta$, dsub&726&2251&42266&0.40 (0.31--0.59)&28.13 (4.71--90.95)\\
3C 236&$d\beta$, dsub&359&1057&4.2/7&0.39 (0.30--1.20)&26.23 (5.84--358.57)\\
3C 285&$d\beta$, dsub&368&1521&4.1/9&0.36 (0.32--0.70)&14.63 (5.98--82.47)\\
3C 293&$d\beta$, dsub&104&2421&1.4/7&0.69 (0.47--1.20)&5.67 (2.29--13.20)\\
3C 296&$d\beta$, ssub&$>$147&6862&6.4/19&0.70 (0.30--1.20)&188.61 (58.28--338.58)\\
3C 303&$s\beta$, dsub&366&2510&3.7/9&0.51 (0.44--0.70)&1.23 (1.00--10.28)\\
3C 305&u/l&&$<$797&&{\it 0.47}&{\it 40.39}\\
3C 310&$d\beta$, ssub&$>$465&9456&126/20&0.50 (0.46--0.54)&41.34 (32.67--53.09)\\
3C 321&$d\beta$, dsub&87&843&6.8/8&1.19 (0.31--1.20)&30.56 (5.64--55.80)\\
3C 326&$d\beta$, dsub&328&321&0.2/5&1.11 (0.30--1.20)&738.42 (10.65--1670.72)\\
3C 338&$d\beta$, ssub&$>$364&363864&1595/39&0.63 (0.61--0.65)&91.37 (85.01--94.86)\\
3C 346&$d\beta$, dsub&411&5079&18/4&0.41 (0.33--0.64)&43.51 (19.53--102.90)\\
3C 386&$s\beta$, dsub&93&384&2.9/3&0.37 (0.30--0.52)&1.01 (1.00--4.17)\\
3C 388&$d\beta$, ssub&$>$667&8149&45/18&0.52 (0.49--0.56)&40.39 (31.04--52.38)\\
3C 390.3&$d\beta$, ssub&$>$294&12647&65/12&0.40 (0.37--0.42)&13.81 (7.33--21.18)\\
3C 433&$d\beta$, dsub&323&3058&42289&1.09 (0.30--1.20)&310.27 (3.75--593.38)\\
3C 442A&$d\beta$, ssub&$>$311&6344&44/14&1.11 (0.58--1.20)&212.86 (101.64--251.01)\\
3C 449&$d\beta$, ssub&$>$119&10260&17/14&0.36 (0.33--0.40)&18.76 (13.11--28.43)\\
3C 452&$d\beta$, dsub&300&3202&7.4/12&0.74 (0.42--1.20)&64.33 (21.33--125.26)\\
3C 465&$d\beta$, ssub&$>$240&37631&18/18&0.53 (0.45--0.74)&155.90 (122.39--216.79)\\
4C 73.08&$s\beta$, dsub&167&624&8.9/4&0.42 (0.31--1.20)&1.01 (1.00--68.33)\\
DA 240&u/l&&$<$433&&{\it 0.47}&{\it 40.89}\\
NGC 6109&$d\beta$, dsub&131&710&4.5/8&1.20 (0.30--1.20)&167.46 (24.80--579.03)\\
NGC 6251&$d\beta$, ssub&$>$220&2084&20/18&0.33 (0.30--0.55)&54.90 (27.11--152.64)\\
NGC 7385&u/l&&$<$1122&&{\it 0.47}&{\it 40.89}\\
PKS 0034-01&u/l&&$<$254&&{\it 0.47}&{\it 40.90}\\
PKS 0038+09&$d\beta$, dsub&231&1238&43/9&0.78 (0.30--1.20)&71.48 (3.96--975.12)\\
PKS 0043-42&$s\beta$, dsub&413&576&7.8/5&0.34 (0.30--0.44)&1.03 (1.00--20.71)\\
PKS 0213-13&$s\beta$, dsub&126&1244&2.9/5&0.63 (0.30--1.20)&18.75 (4.09--81.14)\\
PKS 0349-27&$d\beta$, dsub&310&839&1.0/8&0.30 (0.30--1.50)&97.58 (39.93--1256.53)\\
PKS 0404+03&u/l&&$<$122&&{\it 0.47}&{\it 40.89}\\
PKS 0442-28&u/l&&$<$126&&{\it 0.47}&{\it 40.87}\\
PKS 0620-52&$d\beta$, ssub&$>$368&8843&2.5/10&0.42 (0.37--0.48)&61.59 (41.52--89.35)\\
PKS 0625-35&$d\beta$, ssub&$>$365&5280&19/10&0.36 (0.31--0.67)&40.38 (16.00--168.57)\\
PKS 0625-53&$d\beta$, ssub&$>$412&28390&11/11&0.48 (0.43--0.53)&105.59 (85.04--132.56)\\
PKS 0806-10&u/l&&$<$161&&{\it 0.47}&{\it 40.99}\\
PKS 0915-11&$d\beta$, ssub&$>$315&661760&1339/16&1.20 (0.68--1.20)&302.13 (142.07--417.93)\\
PKS 0945+07&$d\beta$, dsub&318&4404&10/13&0.40 (0.31--0.95)&7.04 (2.04--50.06)\\
PKS 1559+02&$d\beta$, dsub&564&1255&8.7/11&0.30 (0.30--1.20)&6.85 (1.00--1807.76)\\
PKS 1648+05&$d\beta$, ssub&$>$859&25027&538/16&0.63 (0.60--0.65)&93.97 (85.55--101.46)\\
PKS 1733-56&$d\beta$, dsub&492&4055&51/9&1.16 (0.30--1.20)&1165.77 (18.33--1813.91)\\
PKS 1839-48&$d\beta$, dsub&547&6934&21/14&0.66 (0.58--0.75)&191.20 (156.78--239.38)\\
PKS 1949+02&$d\beta$, dsub&252&2435&7.5/8&0.38 (0.30--1.20)&7.36 (1.81--111.99)\\
PKS 1954-55&$d\beta$, ssub&$>$415&7689&25/17&0.36 (0.33--0.41)&30.12 (13.96--51.95)\\
PKS 2211-17&$d\beta$, dsub&784&31037&362/16&0.72 (0.69--0.74)&81.51 (76.04--87.60)\\
PKS 2221-02&$d\beta$, dsub&240&8775&22/16&0.35 (0.30--1.00)&16.56 (7.20--105.59)\\
PKS 2356-61&$d\beta$, dsub&526&1791&1.8/14&0.42 (0.30--1.20)&56.34 (12.69--279.51)\\
\hline\end{tabular}

$^{a}$ Single or double $\beta$ model, or upper limit.
$^{b}$ Single or double background subtraction.
$^{c}$ Upper limits are counts within the median R$_{500}$.
$^{d}$ Counts for {\it XMM-Newton} sources are for the pn camera only.
$^{e}$ Values for $\beta$ and core radius $r_c$ are best fit parameters. Ranges are the Bayesian credible intervals.
$^{f}$ Italics indicate median values used for sources with low counts.
\label{tab:beta}\end{table*}

\begin{table*}
\caption{Radial profile modeling - host galaxy (inner) $\beta$ model}
\begin{tabular}{lcll}\hline
Source&$\chi^2$/dof&$\beta^{a,b}$&r$_c$\\
&&&kpc\\
\hline
3C 28&74/13&0.74 (0.62--0.90)&36.80 (27.05--47.40)\\
3C 31&104/62&0.79 (0.51--1.02)&1.32 (0.32--1.85)\\
3C 33&4.4/6&1.20 (0.85--2.00)&0.99 (0.55--2.01)\\
3C 66B&42351&2.86 (1.09--3.00)&10.19 (4.35--12.24)\\
3C 98&1.4/6&2.92 (0.77--3.00)&0.86 (0.20--1.28)\\
3C 236&4.2/5&1.20 (0.91--2.99)&2.45 (1.60--6.36)\\
3C 285&4.3/7&1.12 (1.00--3.00)&0.79 (0.49--4.11)\\
3C 293&1.6/5&0.98 (0.49--1.20)&0.66 (0.14--1.23)\\
3C 296&7.1/17&0.64 (0.59--0.71)&0.66 (0.48--0.88)\\
3C 310&128/22&0.83 (0.56--2.00)&2.90 (1.67--8.93)\\
3C 321&7.2/6&2.64 (1.07--3.50)&6.12 (2.79--8.33)\\
3C 326&0.2/3&0.95 (0.70--1.50)&10.80 (4.21--35.56)\\
3C 338&1625/37&0.58 (0.56--0.61)&11.94 (11.07--12.86)\\
3C 346&18/5&1.20 (1.20--1.20)&2.54 (1.00--4.44)\\
3C 388&48/16&1.26 (0.79--2.50)&7.47 (4.26--17.71)\\
3C 390.3&65/10&1.07 (0.96--1.25)&5.35 (4.51--6.43)\\
3C 433&12.3/8&1.14 (0.50--1.20)&10.85 (1.49--16.84)\\
3C 442A&44.5/12&0.53 (0.41--0.90)&6.15 (3.01--16.30)\\
3C 449&17.3/12&0.95 (0.72--2.14)&0.96 (0.55--2.21)\\
3C 452&7.5/10&1.52 (0.96--3.00)&1.29 (0.61--2.55)\\
3C 465&20/16&0.66 (0.61--0.73)&0.80 (0.60--1.12)\\
NGC 6109&4.5/6&0.70 (0.47--1.00)&1.21 (0.23--3.26)\\
NGC 6251&18/11&0.71 (0.57--0.88)&0.63 (0.28--1.08)\\
PKS 0038+09&43/7&2.50 (0.96--2.50)&19.46 (8.19--25.83)\\
PKS 0349-27&1.1/6&2.13 (0.45--2.50)&11.05 (1.59--23.59)\\
PKS 0620-52&2.7/8&0.77 (0.61--1.12)&0.73 (0.22--0.97)\\
PKS 0625-35&21/8&0.80 (0.66--0.99)&2.97 (1.93--4.32)\\
PKS 0625-53&3.4/6&0.65 (0.39--0.80)&0.53 (0.11--1.29)\\
PKS 0915-11&237/15&0.89 (0.56--1.44)&30.25 (21.30--40.56)\\
PKS 0945+07&11/11&0.86 (0.71--1.40)&0.60 (0.37--1.52)\\
PKS 1559+02&9.0/9&2.99 (1.14--3.00)&3.25 (1.28--3.75)\\
PKS 1648+05&538/14&2.32 (1.65--3.00)&19.96 (15.00--26.27)\\
PKS 1733-56&51/7&0.56 (0.30--1.27)&2.47 (0.18--10.68)\\
PKS 1839-48&21/12&0.79 (0.50--1.49)&3.03 (0.81--7.17)\\
PKS 1949+02&63/9&0.88 (0.61--1.94)&0.65 (0.36--1.32)\\
PKS 1954-55&14/14&0.61 (0.47--1.14)&2.55 (1.78--8.68)\\
PKS 2211-17&363/14&2.49 (1.58--2.50)&15.23 (10.20--17.05)\\
PKS 2221-02&28/14&1.15 (0.99--1.34)&1.48 (1.16--1.84)\\
PKS 2356-61&1.9/12&1.16 (0.64--2.00)&3.92 (1.14--9.38)\\
\hline\end{tabular}

$^{a}$ Values for $\beta$ and core radius $r_c$ are best fit parameters. Ranges are the Bayesian credible intervals.
$^{b}$ Italics indicate median values used for sources with low counts.
\label{tab:betain}\end{table*}

Luminosity within the $R_{500}$ overdensity radius was calculated by integrating the ICM $\beta$ model profile using counts to flux conversion factors generated from the {\it apec} model. The $\beta$ model profile was extrapolated to $R_{500}$ using the $R-T$ relationship from \citet{arn05}:
\begin{equation}
R_{500}=h(z)^{-1}B_{500}(\frac{kT}{5})^{\beta}
\end{equation}
where
\begin{equation}
h^2(z)=\Omega_{\rm m}(1+z)^3+\Omega_\Lambda
\end{equation}
We calculated a luminosity for each sample of the output of the MCMC code, which provided us with a posterior probability distribution function over luminosity, marginalized over all other parameters. We used the median rather than the mean of the posterior probability distribution function as our luminosity estimate because the distributions were skewed for the fainter sources. Our quoted uncertainties on the luminosity are credible intervals defined on this one-dimensional posterior probability distribution function such that 68~per~cent of the probability is contained in the smallest luminosity range. The luminosity uncertainties take into account the (in some cases large) uncertainties on $\beta$ and $r_c$.

If the cluster emission extended beyond the chip, we could use only the {\it Chandra} blank sky datasets for the background subtraction. The lower energy background varies from field to field and the background is known to change with time \citep{mar03} so the {\it Chandra} background for a given source may not be accurate. We checked how well on average the {\it Chandra} backgrounds modelled the true background by comparing their count rates with those of our source datasets using a chip lying beyond the maximum detected radius of the cluster emission. When the source was centred on the S3 chip, we compared background count rates on the I2 chip, and when it was centred on the I3 chip, we compared background rates on the S2 chip. We then increased the errors on the luminosity using the standard deviation of the differences (14.8~per~cent and 8.5~per~cent respectively).

Luminosity upper limits were calculated within the median $R_{500}$ (533 kpc), using the median $\beta$ and $r_c$ for the profile shapes and the count rate upper limits for the normalisations.

\begin{figure}
  \includegraphics[width=7cm]{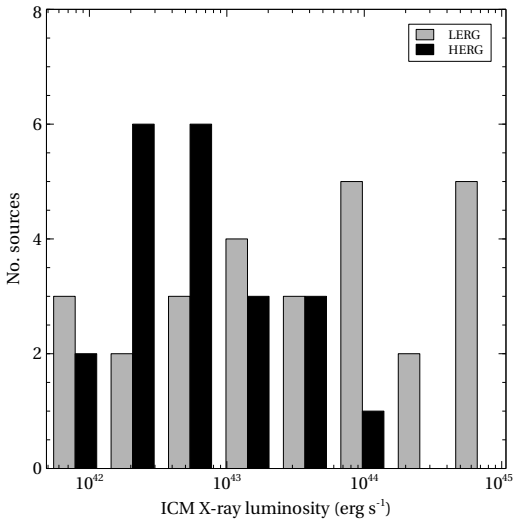}
  \caption{ICM X-ray luminosity distribution for the z0.1 sample, separated into excitation classes (upper limits excluded). Symbols as in Figure~\ref{fig:betas}.}
\label{fig:Lxhisto}\end{figure}

Table~\ref{tab:lum} contains the bolometric X-ray luminosities for each source: within the maximum detection radius; within $R_{500}$; and scaled by $h^{-1}(z)$ to correct for the critical density evolution. Figure~\ref{fig:Lxhisto} shows the luminosity distribution; it can be seen that the most luminous environments are occupied by LERGs.

The sources with emission extending beyond the chip have luminosities from $0.4\times10^{43}$ to $61\times10^{43}$~erg~s$^{-1}$, so cover most of the luminosity range. Since their emission is so extensive, it is to be expected that they tend to be high luminosity. Five of the ten clusters with luminosity greater than $10^{44}$~erg~s$^{-1}$ are among this group, and they include four of the five most luminous clusters. 

\begin{table*}
\caption{ICM X-ray luminosity and electron density}
\begin{tabular}{lllclll}\hline
Source&$L_X (bol)$&$R_{500}$$^{ a}$&$L_X (bol)$&$h(z)^{-1}L_X (bol)$&$n_{e}$\\
&$\times{10}^{43}$~erg~s$^{-1}$&kpc&$\times{10}^{43}$~erg~s$^{-1}$&$\times{10}^{43}$~erg~s$^{-1}$&m$^{-3}$\\
&within $D_{rad}$&&within $R_{500}$&within $R_{500}$&at $0.1R_{500}$\\
\hline
3C 28&$111.300^{+2.329}_{-2.118}$&1175&$107.400^{+1.791}_{-1.753}$&97.558&$3667^{+27}_{-27}$\\
3C 31&$0.626^{+0.056}_{-0.033}$&549&$2.013^{+0.326}_{-0.141}$&1.998&$686^{+30}_{30}$\\
3C 33&$0.489^{+0.017}_{-0.021}$&458&$0.491^{+0.020}_{-0.027}$&0.478&$179^{+84}_{-75}$\\
3C 35&$0.248^{+0.086}_{-0.101}$&420&$0.302^{+0.127}_{-0.174}$&0.293&$772^{+105}_{-94}$\\
3C 66B&$1.444^{+0.030}_{-0.030}$&601&$3.173^{+0.095}_{-0.119}$&3.143&$936^{+10}_{-10}$\\
3C 76.1&$0.042^{+0.020}_{-0.025}$&341&$0.080^{+0.052}_{-0.079}$&0.079&$573^{+163}_{-131}$\\
3C 98&$0.063^{+0.013}_{-0.016}$&330&$0.066^{+0.020}_{-0.021}$&0.065&$407^{+212}_{-197}$\\
3C 192&$0.106^{+0.032}_{-0.035}$&376&$0.155^{+0.057}_{-0.067}$&0.151&$650^{+111}_{-72}$\\
3C 219&$6.866^{+0.869}_{-1.027}$&501&$5.193^{+0.577}_{-0.513}$&4.770&$2708^{+165}_{-159}$\\
3C 236&$0.679^{+0.083}_{-0.111}$&475&$0.693^{+0.095}_{-0.138}$&0.661&$484^{+164}_{-121}$\\
3C 285&$0.522^{+0.115}_{-0.096}$&409&$0.559^{+0.135}_{-0.111}$&0.539&$881^{+62}_{-64}$\\
3C 293&$0.048^{+0.003}_{-0.004}$&373&$0.049^{+0.003}_{-0.005}$&0.048&$290^{+86}_{-122}$\\
3C 296&$0.229^{+0.034}_{-0.034}$&569&$0.567^{+0.128}_{-0.168}$&0.560&$490^{+18}_{-17}$\\
3C 303&$0.749^{+0.113}_{-0.113}$&397&$0.757^{+0.115}_{-0.114}$&0.707&$1547^{+150}_{-107}$\\
3C 305&&{\it 533}&$<$0.078&$<$0.077&$<$201\\
3C 310&$2.997^{+0.448}_{-0.448}$&623&$3.340^{+0.501}_{-0.501}$&3.258&$1940^{+27}_{-27}$\\
3C 321&$0.208^{+0.010}_{-0.010}$&388&$0.211^{+0.011}_{-0.013}$&0.202&$604^{+108}_{-88}$\\
3C 326&$2.000^{+1.182}_{-1.935}$&617&$4.650^{+3.966}_{-4.585}$&4.458&$216^{+28}_{-17}$\\
3C 338&$40.140^{+5.941}_{-5.941}$&1041&$49.440^{+7.319}_{-7.320}$&48.746&$3538^{+4}_{-4}$\\
3C 346&$7.474^{+0.608}_{-0.915}$&731&$10.810^{+1.443}_{-2.280}$&9.991&$2090^{+46}_{-55}$\\
3C 386&$0.033^{+0.007}_{-0.008}$&450&$0.104^{+0.034}_{-0.041}$&0.103&$345^{+39}_{-36}$\\
3C 388&$13.050^{+1.138}_{-1.141}$&866&$14.040^{+1.240}_{-1.230}$&13.450&$2538^{+25}_{-24}$\\
3C 390.3&$3.161^{+0.469}_{-0.469}$&662&$4.262^{+0.637}_{-0.636}$&4.152&$1223^{+23}_{-25}$\\
3C 433&$0.235^{+0.097}_{-0.063}$&409&$0.240^{+0.117}_{-0.069}$&0.229&$485^{+54}_{-168}$\\
3C 442A&$1.129^{+0.101}_{-0.101}$&564&$1.283^{+0.119}_{-0.121}$&1.268&$658^{+17}_{-14}$\\
3C 449&$0.414^{+0.062}_{-0.062}$&583&$1.752^{+0.283}_{-0.280}$&1.738&$911^{+14}_{-15}$\\
3C 452&$0.768^{+0.045}_{-0.054}$&496&$0.788^{+0.052}_{-0.070}$&0.759&$1250^{+66}_{-57}$\\
3C 465&$3.465^{+0.514}_{-0.514}$&1016&$8.903^{+1.541}_{-1.541}$&8.780&$554^{+4}_{-4}$\\
4C 73.08&$0.036^{+0.017}_{-0.025}$&512&$0.049^{+0.029}_{-0.042}$&0.048&$258^{+154}_{-90}$\\
DA 240&&{\it 533}&$<$0.083&$<$0.081&$<$290\\
NGC 6109&$0.136^{+0.036}_{-0.052}$&412&$0.239^{+0.105}_{-0.154}$&0.236&$577^{+41}_{-41}$\\
NGC 6251&$0.183^{+0.027}_{-0.027}$&523&$0.444^{+0.074}_{-0.071}$&0.439&$348^{+17}_{-16}$\\
NGC 7385&&{\it 533}&$<$0.064&$<$0.063&$<$256\\
PKS 0034-01&&{\it 533}&$<$0.063&$<$0.061&$<$264\\
PKS 0038+09&$2.418^{+0.458}_{-0.446}$&566&$2.652^{+0.596}_{-0.776}$&2.418&$1754^{+504}_{-354}$\\
PKS 0043-42&$1.018^{+0.198}_{-0.275}$&543&$1.317^{+0.282}_{-0.397}$&1.246&$863^{+61}_{-55}$\\
PKS 0213-13&$0.184^{+0.060}_{-0.075}$&375&$0.205^{+0.080}_{-0.101}$&0.190&$1193^{+316}_{-237}$\\
PKS 0349-27&$0.280^{+0.132}_{-0.189}$&392&$0.332^{+0.176}_{-0.246}$&0.322&$270^{+36}_{-69}$\\
PKS 0404+03&&{\it 533}&$<$0.268&$<$0.257&$<$549\\
PKS 0442-28&&{\it 533}&$<$0.273&$<$0.254&$<$304\\
PKS 0620-52&$3.941^{+0.590}_{-0.589}$&768&$6.742^{+1.047}_{-1.036}$&6.585&$1261^{+18}_{-18}$\\
PKS 0625-35&$3.150^{+0.477}_{-0.477}$&872&$6.334^{+1.067}_{-1.053}$&6.176&$883^{+24}_{-27}$\\
PKS 0625-53&$22.260^{+1.912}_{-1.917}$&1287&$42.610^{+3.928}_{-3.945}$&41.560&$1609^{+16}_{-15}$\\
PKS 0806-10&&{\it 533}&$<$0.206&$<$0.195&$<$482\\
PKS 0915-11&$54.560^{+8.075}_{-8.075}$&851&$61.800^{+9.146}_{-9.147}$&60.248&$5140^{+8}_{-11}$\\
PKS 0945+07&$1.608^{+0.128}_{-0.173}$&562&$1.790^{+0.206}_{-0.313}$&1.719&$903^{+89}_{-94}$\\
PKS 1559+02&$1.041^{+0.233}_{-0.441}$&329&$0.790^{+0.160}_{-0.180}$&0.752&$803^{+348}_{-110}$\\
PKS 1648+05&$57.830^{+8.571}_{-8.570}$&995&$58.620^{+8.689}_{-8.687}$&54.375&$4976^{+24}_{29}$\\
PKS 1733-56&$0.826^{+0.349}_{-0.604}$&506&$0.842^{+0.365}_{-0.618}$&0.803&$680^{+35}_{-51}$\\
PKS 1839-48&$18.730^{+0.391}_{-0.445}$&1156&$23.640^{+0.665}_{-0.825}$&22.424&$1570^{+26}_{-24}$\\
PKS 1949+02&$0.217^{+0.041}_{-0.061}$&411&$0.259^{+0.068}_{-0.100}$&0.252&$610^{+65}_{-70}$\\
PKS 1954-55&$4.496^{+0.109}_{-0.109}$&785&$7.767^{+0.323}_{-0.310}$&7.560&$1262^{+23}_{-21}$\\
PKS 2211-17&$28.360^{+0.232}_{-0.262}$&916&$28.640^{+0.238}_{-0.265}$&26.593&$4397^{+23}_{-26}$\\
PKS 2221-02&$1.169^{+0.056}_{-0.086}$&450&$1.428^{+0.128}_{-0.222}$&1.391&$782^{+34}_{-38}$\\
PKS 2356-61&$1.452^{+0.308}_{-0.424}$&465&$1.383^{+0.252}_{-0.373}$&1.321&$1049^{+152}_{-120}$\\
\hline\end{tabular}

$^{a}$ Italics indicate median value used for sources with low counts.
\label{tab:lum}\end{table*}

\begin{figure}
  \begin{minipage}{8cm}
  \includegraphics[width=7cm]{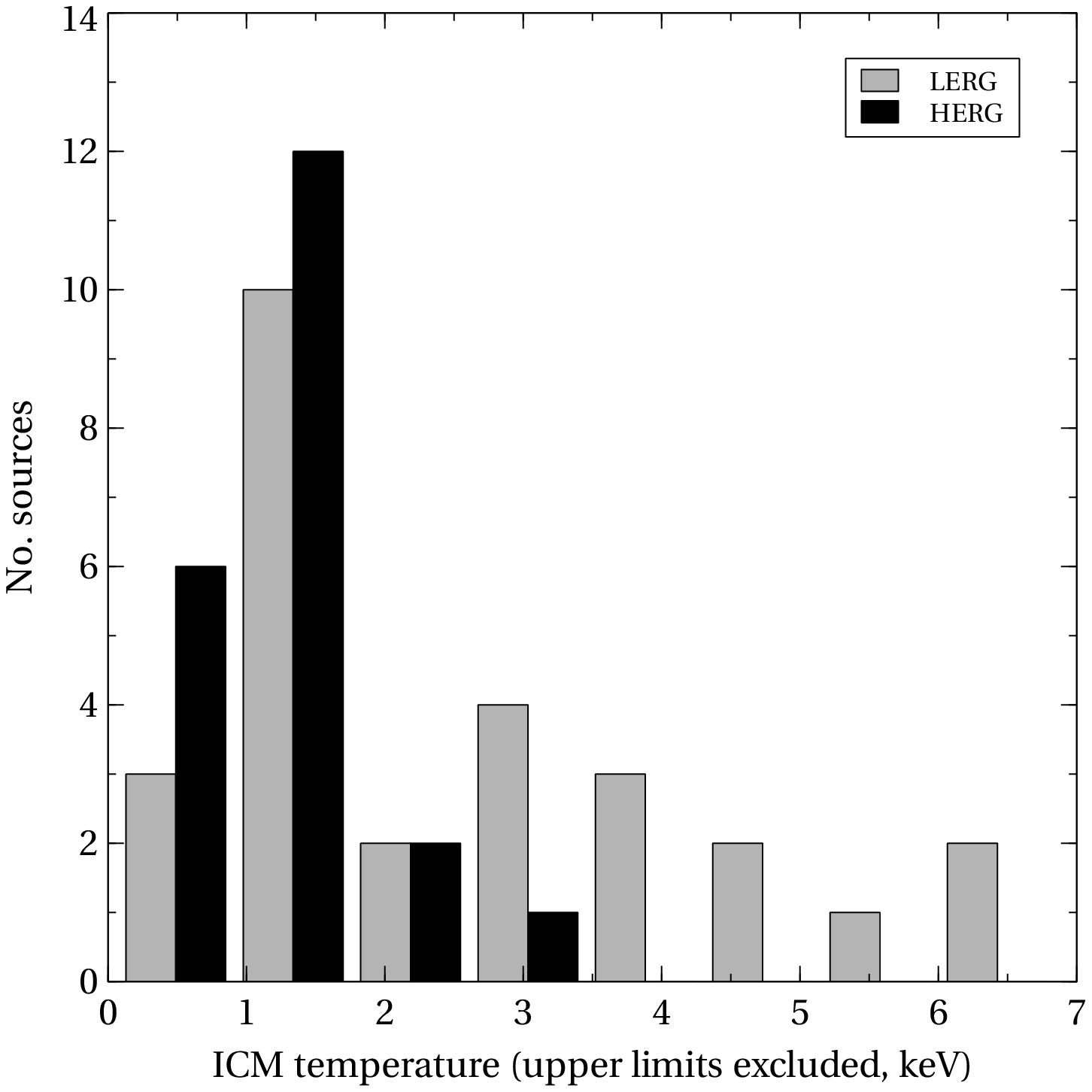}
  \end{minipage}
  \begin{minipage}{8cm} 
  \includegraphics[width=7cm]{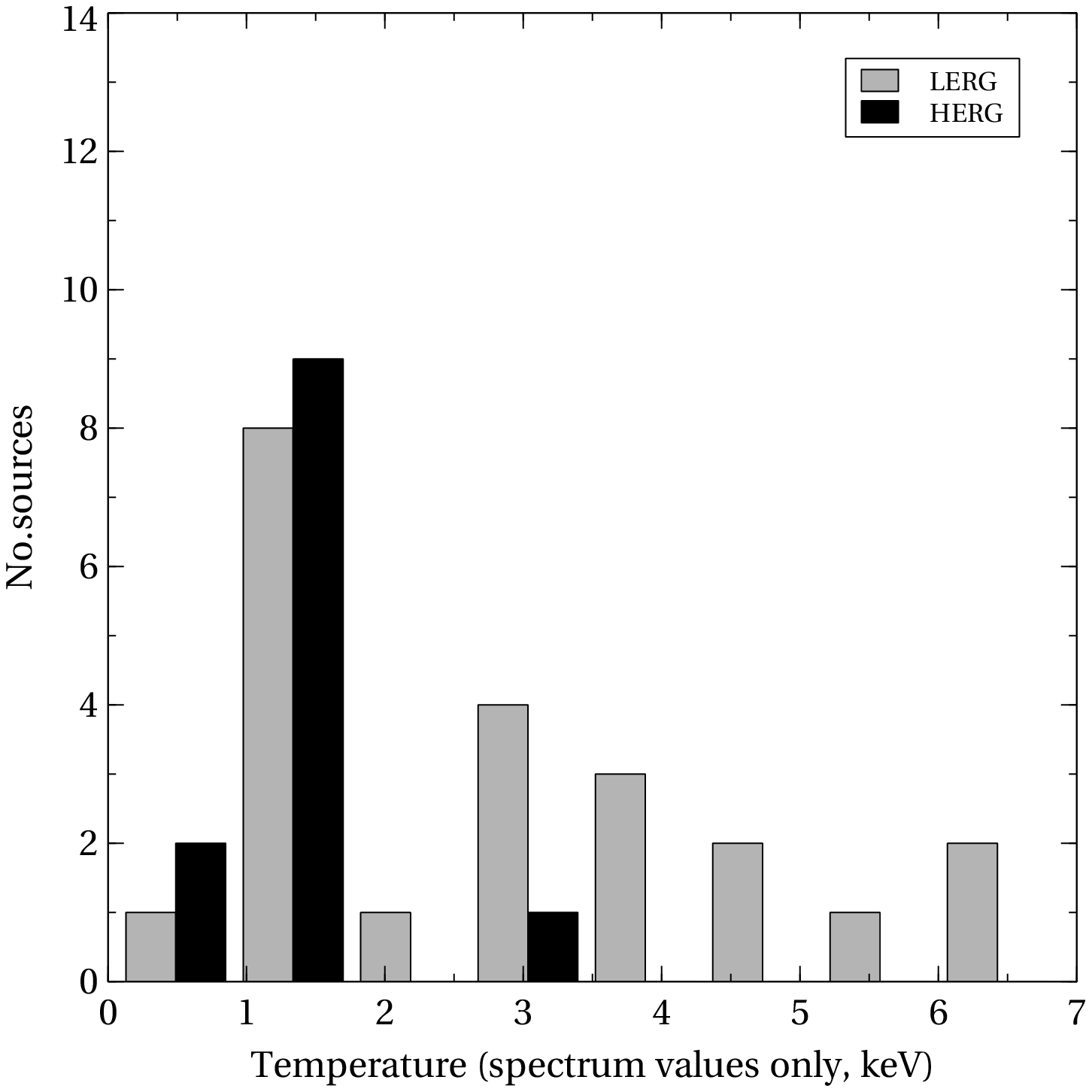}
  \end{minipage}
  \caption{ICM temperature distribution for the z0.1 sample, separated into excitation classes (upper limits excluded). The upper histogram includes all sources; the lower histogram shows only those with temperatures obtained by spectral analysis. Symbols as in Figure~\ref{fig:betas}.}
\label{fig:Txhisto}\end{figure}

We also wished to gain some idea of the central conditions of the radio galaxy environments so we calculated the environment density at a radius of $0.1R_{500}$; for all but two of sources this was not much larger than the host galaxy radius and so was the closest radius to the source that we could measure the density. We used the method described in \citet{bir93}:
\begin{equation}
\mathscr{A}=\displaystyle\frac{\int{n_{e}n_{p}dV}}{4\pi d\Omega D_{L}^{2}}
\end{equation}
\noindent where $\mathscr{A}$ is the distance-normalised volume emission measure of the atmosphere per unit solid angle, $n_{e}$ and $n_{p}$ are the electron and proton densities, $dV$ is a cylindrical volume element subtending a solid angle $d\Omega$ to the observer and $D_{L}$ is the luminosity distance to the source. We assumed an electron-proton ratio of 1.18.

We obtained the emission measure normalisation from the normalisation of the \textit{apec} model used in the spectral analysis (see below). We then calculated the electron density at $0.1R_{500}$ for each sample in the MCMC code output, using the same method that we used for the luminosity calculations (see above). The uncertainties were also derived in the same manner. The densities are included in Table~\ref{tab:lum}.

\subsection{Spectral analysis}

When there were sufficient counts, we obtained the ICM temperature from spectral analysis. We used the \textsc{xspec} package, using the \textit{apec} model for the thermal bremsstrahlung from the ICM and the \textit{wabs} photo-electric absorption model to take account of Galactic absorption. We reduced the energy range to 0.5 to 5.0 keV to exclude the PSF-scattered high energy emission from the nucleus and, where possible, we used a double subtraction method similar to that used for the {\it XMM-Newton} temperatures in I13, using the following steps:

\begin{enumerate}[i]
\item Generate the spectrum for a background region beyond the maximum detected radius with {\it specextract}, using the {\it Chandra} blank sky files or {\it XMM-Newton} closed filter files for the spectrum background file. Then scale the blank sky background spectrum to match the source file count rate (see Section~\ref{sec:Xray});
\item Fit models to this background region using a power law with index 1.41 to model the extragalactic component of the X-ray background \citep{lum02}, with Galactic absorption (\textit{wabs(power)}). We added local thermal models (\textit{apec}) for the Milky Way emission and local hot bubble as appropriate;
\item Generate the spectrum for the ICM using an annulus from where the ICM dominates over the PSF and host galaxy emission out to the maximum detection radius. Again, the {\it Chandra} blank sky files or {\it XMM-Newton} closed filter files are used for the spectrum background and the blank sky spectrum is scaled to match the source file count rate;
\item Scale the normalisations of the background model elements to match the source region area;
\item Model the source bremsstrahlung using \textit{wabs(apec)}, with the elements of the background model as fixed parameters. When possible, we left the metallicity as a free parameter. If it did not converge to a reasonable value (0.15 to 0.6~solar -- \citealt{bal07}), or had errors larger than this range, we used 0.3~solar.
\end{enumerate}

In some sources, the {\it Chandra} blank sky files matched the background from the source observation sufficiently well that there were very few counts left after the background subtraction. In these cases, we used the same method as I13, using single subtraction with a region from outside the maximum detected radius as the background for the spectrum.

If the source emission was so extensive that there was no region of the source observation that could be used for background, we could not use double subtraction. In this case, single subtraction using the blank sky files as background was the only method available. In addition to the ICM emission, the cosmic ray background and local thermal emission were modelled as in the double subtraction method, but were left as free parameters.

When there were sufficient counts, we generated a temperature profile and selected an annulus for the ICM temperature across a wide range of stable results, excluding the cool core if present. Otherwise we obtained temperatures for a range of annuli, looking for the lowest $\chi^2$.

For 3C~31 and 3C~66B, we also excluded the 1.4 to 1.6 keV energy band from the spectrum to remove the  \textit{XMM-Newton} 1.5 keV instrumental aluminium fluorescence line \citep{fre02}.

The spectra for 18 sources had insufficient counts for modelling the spectrum. For these sources we extracted a spectrum to the maximum detected radius and estimated the temperature in the same way as in I13, using the counts from the $\beta$ model to estimate luminosity from the \textit{apec} model, and then using the scaling relation of \citet{pra09} to estimate the temperature. This was used to estimate $R_{500}$, and the process was iterated until we obtained a stable temperature. We used the same process for the seven sources without a $\beta$ model fit, using the 3$\sigma$ upper limit on the counts to estimate the starting luminosity.

The distribution of temperatures is shown in Figure~\ref{fig:Txhisto}, and Table~\ref{tab:temp} contains the inner and outer radii of the annuli, the temperatures of the sources and the $\chi^2$ for the temperatures obtained by spectral analysis. The temperatures range from 0.65 to 6.8 keV -- a similar range to the ERA sample -- and are for the most part typical of groups and poor clusters. The temperatures for the sources where the emission extended beyond the chip included the highest temperature (6.8 keV), but otherwise ranged from 1.38 keV to 4.43 keV, not concentrated in or dominating any temperature range. Although both the distributions of both types of radio galaxy peak at low temperatures, the LERGs have a wider range of temperatures with the top of the range being occupied exclusively by LERGs.

\begin{table*}
\caption{ICM temperatures}
\begin{tabular}{llcclll}\hline
Source&Method&Annulus radii&Metallicity$^{a}$&Temperature&$\chi^2/dof$$^{b}$\\
&&arcsec&$Z/Z_{\odot}$&keV&\\
\hline
3C 28&Spectrum&24.60-246&0.30&$6.61^{+0.59}_{-0.49}$&106/143\\
3C 31&Spectrum&60.00-600&0.30&$1.52^{+0.03}_{-0.03}$&2198/1430\\
3C 33&Estimate&2.46-98&0.30&$1.12^{+0.01}_{-0.02}$& \\
3C 35&Estimate&2.46-172&0.30&$0.97^{+0.10}_{-0.20}$& \\
3C 66B&Spectrum&50.00-600&$0.21^{+0.04}_{-0.03}$&$1.71^{+0.05}_{-0.04}$&711/691\\
3C 76.1&Estimate&2.46-147&0.30&$0.80^{+0.07}_{-0.12}$& \\
3C 98&Estimate&2.95-196&0.30&$0.62^{+0.04}_{-0.06}$& \\
3C 192&Estimate&1.48-147&0.30&$0.80^{+0.07}_{-0.12}$& \\
3C 219&Spectrum&4.92-172&0.15&$1.46^{+0.13}_{-0.14}$&36/28\\
3C 236&Estimate&2.46-196&0.30&$1.24^{+0.05}_{-0.08}$& \\
3C 285&Spectrum&34.40-172&0.30&$0.94^{+0.10}_{-0.22}$&22/17\\
3C 293&Spectrum&2.46-73&0.30&$0.78^{+0.06}_{-0.06}$&18/12\\
3C 296&Spectrum&19.68-147&0.15&$1.59^{+0.12}_{-0.12}$&45/54\\
3C 303&Spectrum&2.46-123&0.30&$0.94^{+0.09}_{-0.13}$&7.2/7\\
3C 305&Upper limit&9.84-147&0.30&$<0.65$&\\
3C 310&Spectrum&9.84-246&$0.26^{+0.07}_{-0.06}$&$1.92^{+0.12}_{-0.13}$&135/134\\
3C 321&Estimate&2.46-49&0.30&$0.87^{+0.01}_{-0.02}$& \\
3C 326&Estimate&14.76-196&0.30&$1.94^{+0.40}_{-1.35}$& \\
3C 338&Spectrum&19.68-344&$0.44^{+0.04}_{-0.04}$&$4.63^{+0.08}_{-0.08}$&392/304\\
3C 346&Spectrum&2.46-98&0.30&$2.79^{+0.24}_{-0.21}$&111/94\\
3C 386&Spectrum&2.46-270&0.30&$1.05^{+0.18}_{-0.12}$&12/15\\
3C 388&Spectrum&9.84-344&$0.53^{+0.18}_{-0.15}$&$3.52^{+0.18}_{-0.15}$&141/167\\
3C 390.3&Estimate&2.46-270&0.30&$2.14^{+0.01}_{-0.01}$& \\
3C 433&Spectrum&9.84-147&0.30&$0.96^{+0.76}_{-0.27}$&6.8/10\\
3C 442A&Spectrum&14.76-319&$0.24^{+0.12}_{-0.09}$&$1.58^{+0.11}_{-0.17}$&75/85\\
3C 449&Spectrum&14.76-147&$0.33^{+0.09}_{-0.08}$&$1.66^{+0.06}_{-0.07}$&63/77\\
3C 452&Spectrum&9.84-147&0.30&$1.32^{+0.10}_{-0.08}$&45/46\\
3C 465&Spectrum&9.84-295&$0.36^{+0.09}_{-0.08}$&$4.43^{+0.26}_{-0.23}$&255/259\\
4C 73.08&Spectrum&14.76-147&0.30&$1.37^{+0.26}_{-0.18}$&8.3/8\\
DA 240&Upper limit&14.76-147&0.30&$<0.66$&\\
NGC 6109&Estimate&9.84-221&0.30&$0.91^{+0.11}_{-0.24}$& \\
NGC 6251&Spectrum&34.44-246&0.30&$1.38^{+0.21}_{-0.03}$&30/25\\
NGC 7385&Upper limit&4.92-147&0.30&$<0.61$&\\
PKS 0034-01&Upper limit&9.84-147&0.30&$<0.61$&\\
PKS 0038+09&Estimate&2.46-73&0.30&$1.82^{+0.12}_{-0.18}$& \\
PKS 0043-42&Spectrum&2.46-196&$0.41^{+0.69}_{-0.69}$&$1.59^{+0.96}_{-0.33}$&1.2/6\\
PKS 0213-13&Estimate&4.92-49&0.30&$0.85^{+0.09}_{-0.15}$& \\
PKS 0349-27&Spectrum&2.95-246&0.30&$0.86^{+0.16}_{-0.20}$&15/13\\
PKS 0404+03&Upper limit&19.68-147&0.30&$<0.93$&\\
PKS 0442-28&Upper limit&9.84-147&0.30&$<0.93$&\\
PKS 0620-52&Spectrum&19.68-246&$0.46^{+0.41}_{-0.19}$&$2.76^{+1.10}_{-0.60}$&115/137\\
PKS 0625-35&Spectrum&4.92-246&0.30&$3.46^{+0.48}_{-0.31}$&64/83\\
PKS 0625-53&Spectrum&24.60-295&0.30&$6.84^{+0.38}_{-0.34}$&171/188\\
PKS 0806-10&Upper limit&19.68-147&0.30&$<0.86$&\\
PKS 0915-11&Spectrum&2.46-246&$0.30^{+0.02}_{-0.02}$&$3.31^{+0.10}_{-0.11}$&399/301\\
PKS 0945+07&Estimate&2.46-196&0.30&$1.64^{+0.06}_{-0.10}$& \\
PKS 1559+02&Spectrum&4.92-221&0.30&$0.65^{+0.10}_{-0.18}$&15/25\\
PKS 1648+05&Spectrum&14.76-196&$0.24^{+0.11}_{-0.10}$&$4.29^{+0.11}_{-0.11}$&200/239\\
PKS 1733-56&Spectrum&9.84-196&0.30&$1.38^{+0.42}_{-0.08}$&22/20\\
PKS 1839-48&Spectrum&4.92-147&0.30&$5.95^{+0.80}_{-0.72}$&89/102\\
PKS 1949+02&Estimate&4.92-221&0.30&$0.93^{+0.07}_{-0.12}$& \\
PKS 1954-55&Spectrum&4.92-246&0.30&$2.89^{+0.22}_{-0.22}$&172/121\\
PKS 2211-17&Spectrum&24.60-196&0.30&$4.11^{+0.18}_{-0.18}$&165/134\\
PKS 2221-02&Spectrum&4.92-221&0.30&$1.09^{+0.60}_{-0.40}$&52/57\\
PKS 2356-61&Spectrum&4.92-196&0.30&$1.19^{+0.14}_{-0.20}$&17/15\\
\hline\end{tabular}

$^{a}$ Metallicities without errors were used as fixed parameters.
$^{b}$ For temperatures from spectral analysis.
\label{tab:temp}\end{table*}

\section{RESULTS AND DISCUSSION}

\subsection{Radio galaxy environments}\label{sec:LrLx}

\subsubsection{z0.1 sample results}

\begin{figure*}
  \begin{minipage}{8cm}
  \includegraphics[width=7cm]{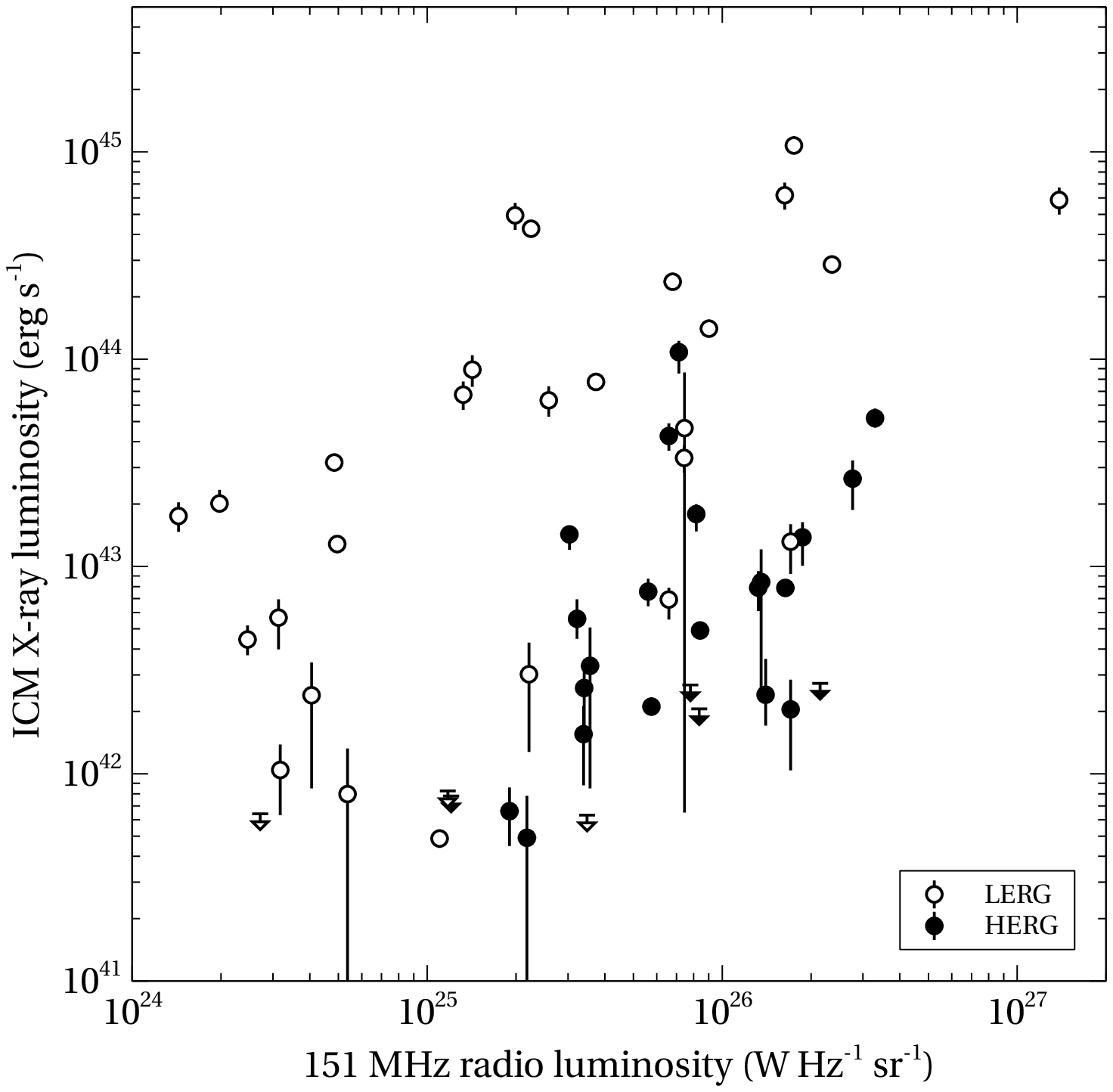}
  \end{minipage}
  \begin{minipage}{8cm} 
  \includegraphics[width=7cm]{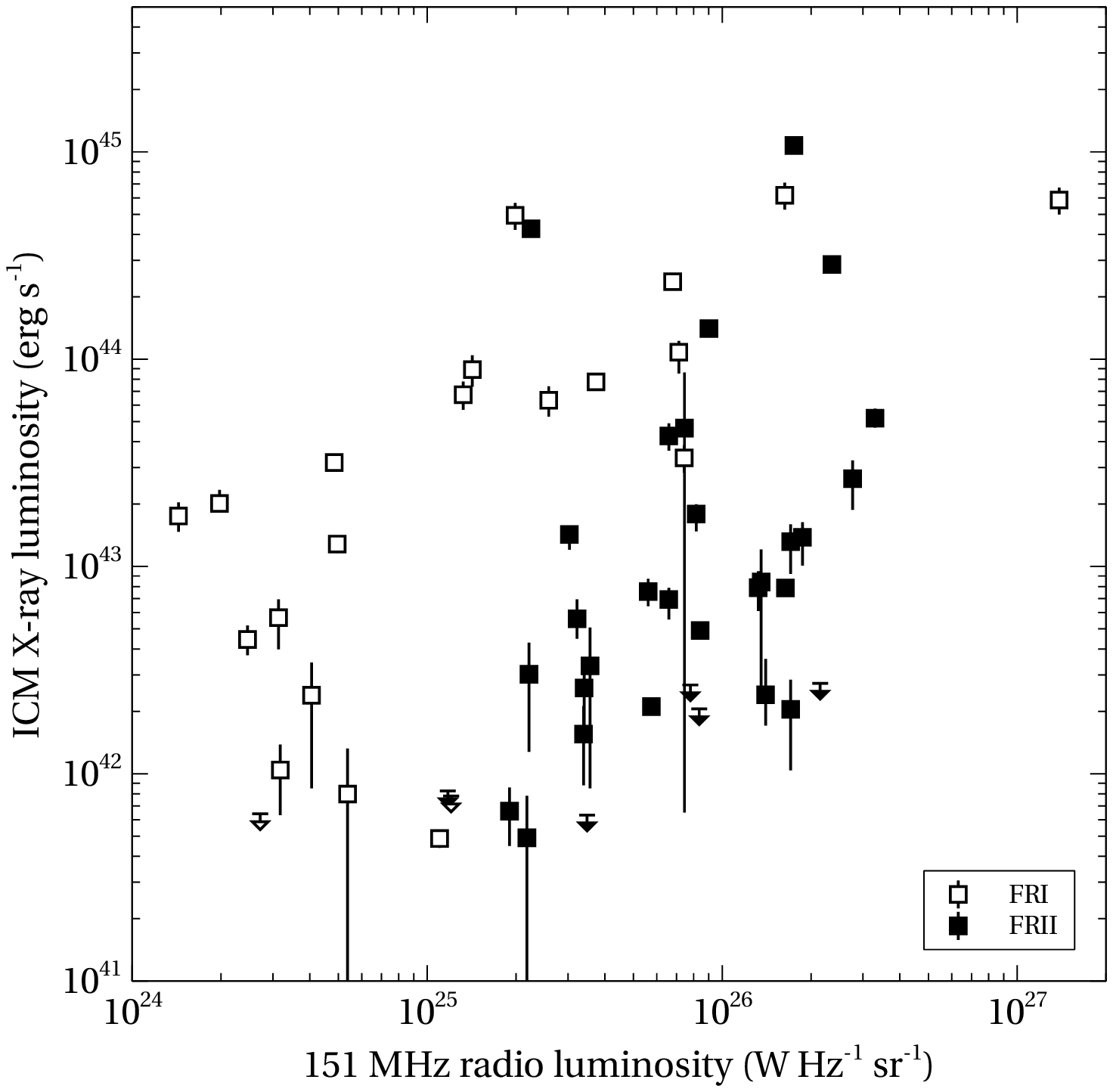}
  \end{minipage}
  \caption{Radio luminosity vs ICM X-ray luminosity for the z0.1 sample, separated into excitation classes (left), and FRI and FRII galaxies (right). LERGs/FRIs are shown by empty symbols and HERGs/FRIIs by filled symbols. Corresponding upper limits are shown by empty and filled arrows.}
\label{fig:LrLx}\end{figure*}

We first compared radio and ICM luminosities for the z0.1 sample, to see if the results differed from those found for the ERA sample. As can be seen from Figure~\ref{fig:LrLx} (left), the HERG subsample appears to cluster in the lower, central region of the plot, while the LERG sub-sample forms a diagonal across the plot. Since there is a strong correlation between radio luminosity and redshift from the Malmquist bias and the lack of high luminosity local sources (see Figure~\ref{fig:zLr}), we used partial correlation Kendall's $\tau$ tests \citep{aks96} to look for correlations between the radio and ICM luminosities in the presence of a dependency on redshift, and found a weak correlation for the full sample, a strong correlation for the LERG sub-sample ($>3\sigma$) and no correlation for the HERGs (Table~\ref{tab:KTau}). These results agree with those found with the ERA sample at $z\sim0.5$, and, as discussed in Section~\ref{sec:intro}, by \citet{bes04,har04,bel07,gen13}.

Because the sources are relatively close, the maximum detected radius is on average only half of $R_{500}$. Consequently, the beta models needed to be extrapolated to calculate the luminosities for all but ten of the sources. We therefore also calculated the luminosity for each source within $0.5R_{500}$ (which was within the maximum detected radius for 80~per~cent of the sources) to check that the statistical results were similar. This gave similar results for the partial correlation tests between the sub-samples ($>3\sigma$ for the LERGS, no correlation for the HERGs), implying that the extrapolation to $R_{500}$ does not have a significant effect.

We also checked the LERG results for a sub-sample with $z>0.03$, removing the bulk of the redshift dependence, which gave a weaker correlation ($>2\sigma$). This was to be expected, since the scatter is proportionally greater over the reduced luminosity range. 

For the FRI and FRII sub-samples in Figure~\ref{fig:LrLx} (right), there appears to be a correlation between radio and ICM luminosity for both sub-samples, and this is confirmed by the partial correlation tests (Table~\ref{tab:KTau}) ($>2.5\sigma$ for both sub-samples). Since the FRI and FRII sub-samples contain both HERGs and LERGs, one would expect their correlation strengths to lie between those of the HERGs and LERGs, and this is the case.

\begin{table*}
\caption{Partial correlation analysis results, using Generalized Kendall's $\tau$ correlation tests in the presence of a correlation with a third factor.}
\begin{tabular}{llccc}\hline
Sample&Sub-sample&N&$\tau/\sigma$&p\\
\hline
\multicolumn{4}{|l|}{Radio luminosity vs ICM luminosity, with a redshift correlation}\\
z0.1&All&55&2.62&0.0088\\
 &HERG&25&1.68&0.0930\\
&LERG&30&3.39&0.0006\\
&FRI&22&2.96&0.0030\\
&FRII&33&2.91&0.0036\\
\multicolumn{4}{|l|}{Radio luminosity vs ICM luminosity, with a redshift correlation}\\
0.03$<$z$<$0.2&All&46&2.37&0.0178\\
&HERG&25&1.68&0.0930\\
&LERG&21&2.72&0.0066\\
\multicolumn{4}{|l|}{Radio luminosity vs ICM luminosity, with a redshift correlation}\\
z0.1 and ERA&All&81&4.09&0.0001\\
&HERG&40&2.43&0.0150\\
&LERG&41&5.08&$<0.0001$\\
\multicolumn{4}{|l|}{Radio luminosity vs ICM luminosity, with a redshift correlation}\\
Matched&HERG&28&0.79&0.4296\\
z0.1 and ERA&LERG&27&4.69&$<0.0001$\\
\multicolumn{4}{|l|}{ICM luminosity vs temperature, with a redshift correlation}\\
z0.1&Spectrum&34&15.28&$<0.0001$\\
z0.1 and ERA&Spectrum&44&18.35&$<0.0001$\\
\multicolumn{4}{|l|}{ICM luminosity vs $B_{gq}$, with a redshift correlation}\\
All&All&22&3.66&0.0002\\
\multicolumn{4}{|l|}{Radio luminosity vs central density, with a redshift correlation}\\
z0.1&HERG&25&0.73&0.4654\\
&LERG&30&2.84&0.0046\\
\multicolumn{4}{|l|}{ICM luminosity vs central density, with a redshift correlation}\\
&HERG&25&5.09&$<0.0001$\\
&LERG&30&7.79&$<0.0001$\\
\multicolumn{4}{|l|}{Radio luminosity vs ICM luminosity, with a central density correlation}\\
&HERG&25&1.32&0.1868\\
&LERG&30&1.50&0.1336\\
\multicolumn{4}{|l|}{Radio luminosity vs central density, with an ICM luminosity correlation}\\
&HERG&25&1.14&0.2542\\
&LERG&30&1.37&0.1706\\
\hline\end{tabular}

N is sample size; $\tau$ is the partial correlation statistic; $\sigma$ is the standard deviation; p is probability under the null hypothesis.
\label{tab:KTau}\end{table*}

\subsubsection{Combined z0.1 and ERA samples}\label{sec:LrLxcomb}

\begin{figure*}
  \begin{minipage}{8cm}
  \includegraphics[width=7cm]{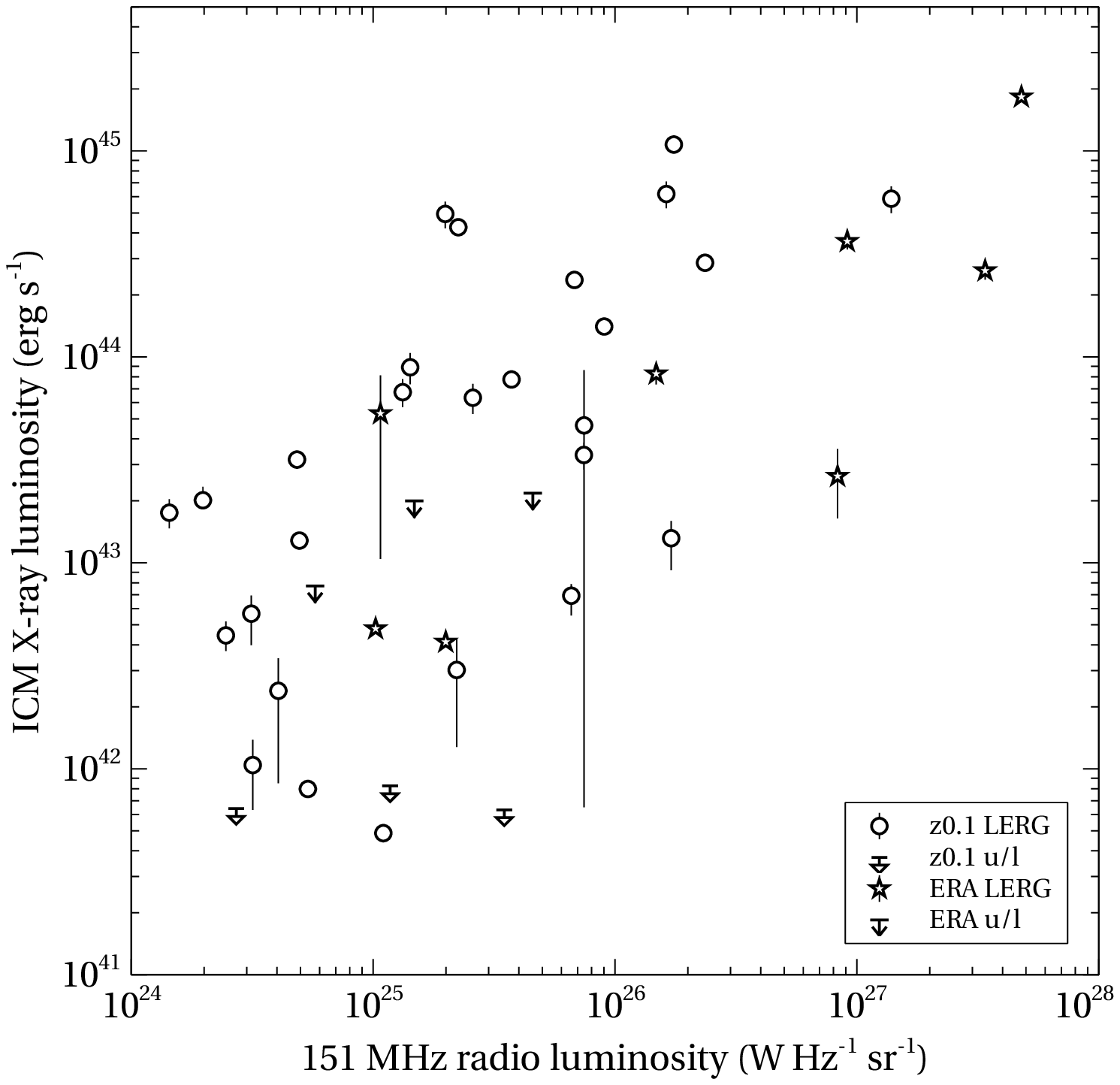}
  \end{minipage}
  \begin{minipage}{8cm} 
  \includegraphics[width=7cm]{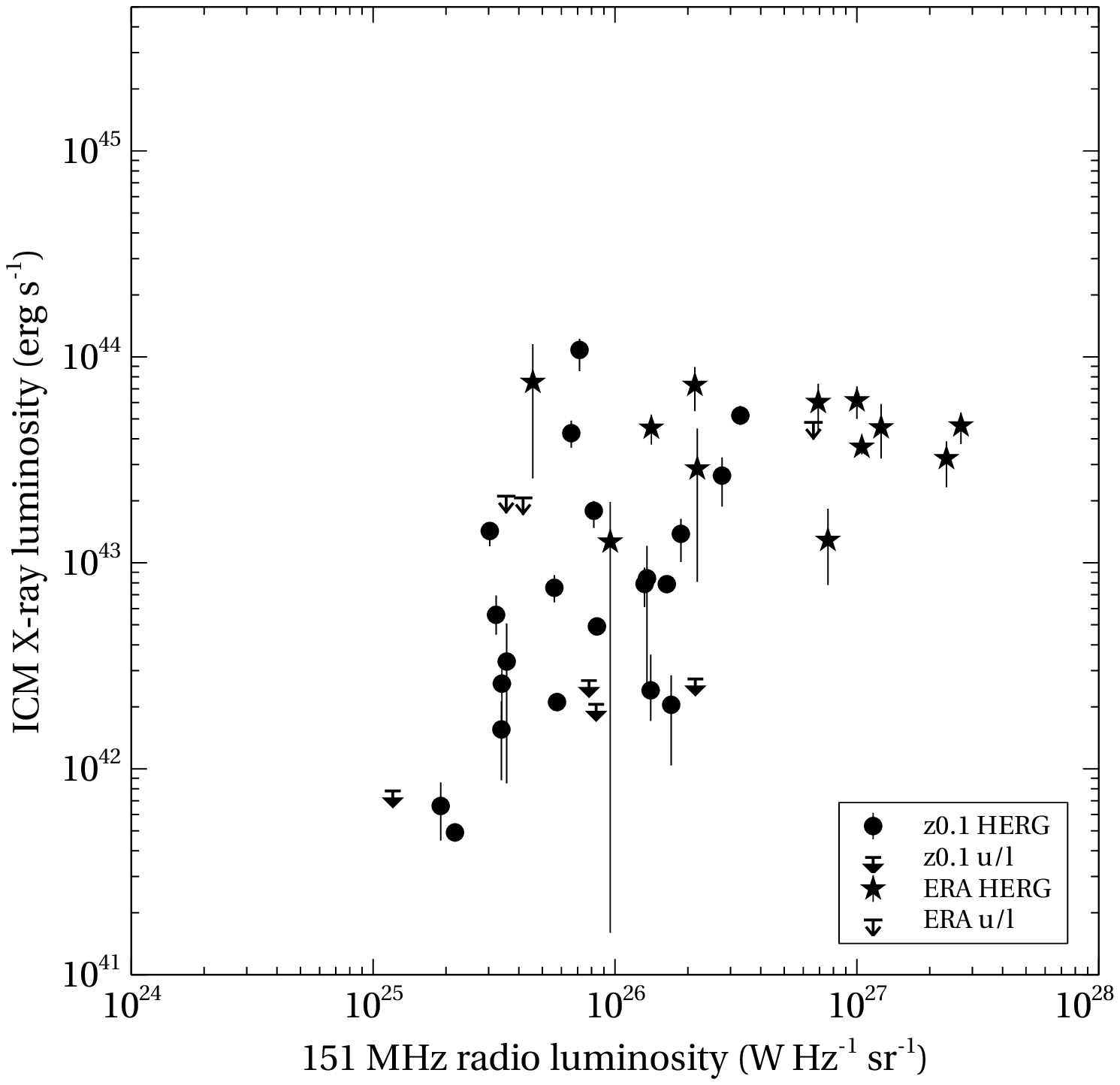}
  \end{minipage}
  \caption{Radio luminosity vs ICM X-ray luminosity for the combined z0.1 and ERA samples, with LERGs on the left (empty symbols) and HERGs on the right (filled symbols). z0.1 sources are shown as circles and ERA sources as stars. z0.1 upper limits use triangles and ERA upper limits use arrows.}
\label{fig:LrLxAll}\end{figure*}

Figure~\ref{fig:LrLxAll} shows the radio vs ICM luminosities for the combined ERA and $z \sim 0.1$ samples for the LERGs (left) and HERGs (right). Both of the LERG samples occupy the same diagonal across the plot, with a similar amount of scatter, and the partial correlation test gives a strong correlation ($>4\sigma$). The HERGs show a much weaker correlation ($>2\sigma$) -- looking at Figure~\ref{fig:LrLxAll} the correlation appears to come from the absence of low ICM luminosities for the ERA sources. See Section~\ref{sec:Hevol} below for a discussion of the lack of sources in this region.

The z0.1 and ERA samples cover different ranges of radio luminosity. We therefore checked our result using sub-samples matched in luminosity ranges: $5\times10^{24}-5\times10^{26}$~W~Hz$^{-1}$~sr$^{-1}$ for the LERGs and $3\times10^{25}-3\times10^{26}$~W~Hz$^{-1}$~sr$^{-1}$ for the HERGs (Table~\ref{tab:KTau}). In these sub-samples, the strong LERG $L_X-L_R$ correlation is still present, but since the high radio luminosity HERGs are no longer in the sample, the partial correlation Kendall's $\tau$ test now showed no correlation for the HERGs. The reduced ERA samples are small and contain upper limits on the ICM luminosities so may affect the accuracy of the statistics; nevertheless, the difference between the HERG and LERG results in both the complete and matched samples is striking.

\subsubsection{Redshift evolution}\label{sec:Hevol}

\begin{figure*}
  \begin{minipage}{8cm}
  \includegraphics[width=7cm]{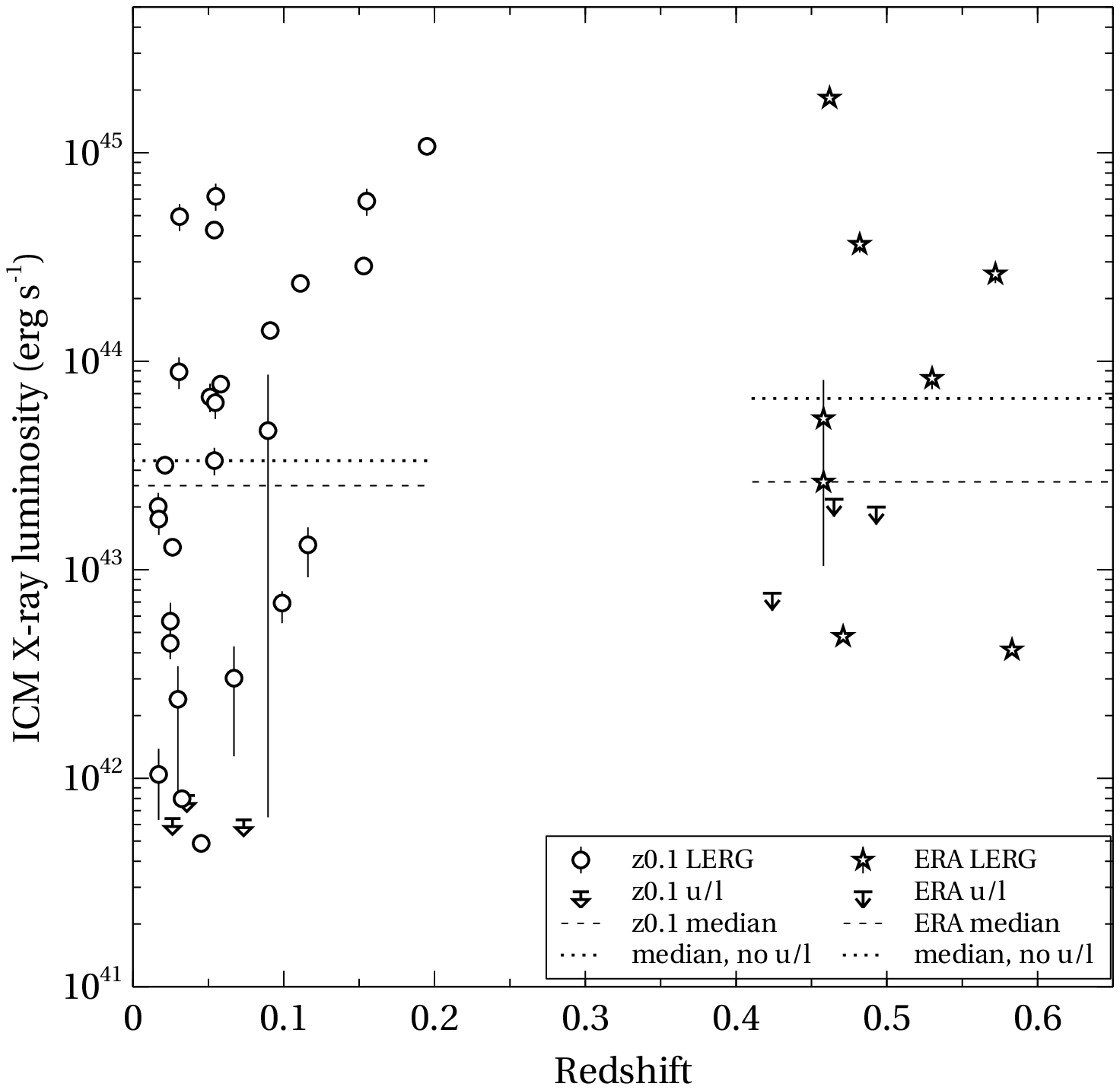}
  \end{minipage}
  \begin{minipage}{8cm} 
  \includegraphics[width=7cm]{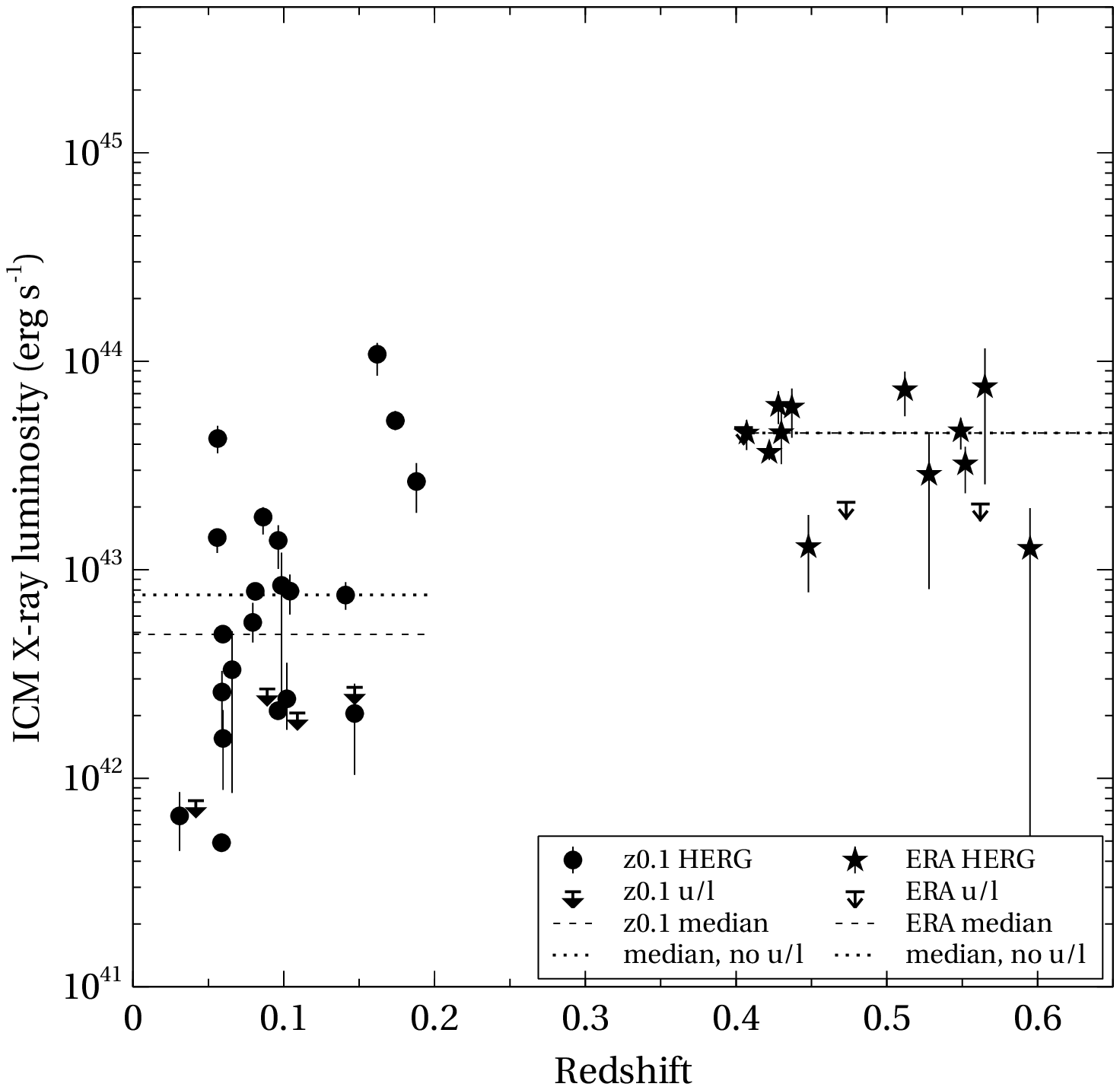}
  \end{minipage}
  \caption{ICM X-ray luminosity vs redshift for the combined z0.1 and ERA samples, with LERGs on the left and HERGs on the right. Symbols as in Figure~\ref{fig:LrLxAll}.}
\label{fig:zLx}\end{figure*}

Looking at Figure~\ref{fig:zLx}, there seems to be little difference between the LERG samples from the two epochs, suggesting that there has been no evolution of the environment since $z\sim0.5$. The ERA HERG sample, however, occupies a narrower range of ICM luminosities than the z0.1 HERG sample, but with similar maximum values. This may indicate that there has been evolution of the HERG environments, as suggested by \citet{harv01} and \citet{bel07}.

\begin{figure*}
  \begin{minipage}{8cm}
  \includegraphics[width=7cm]{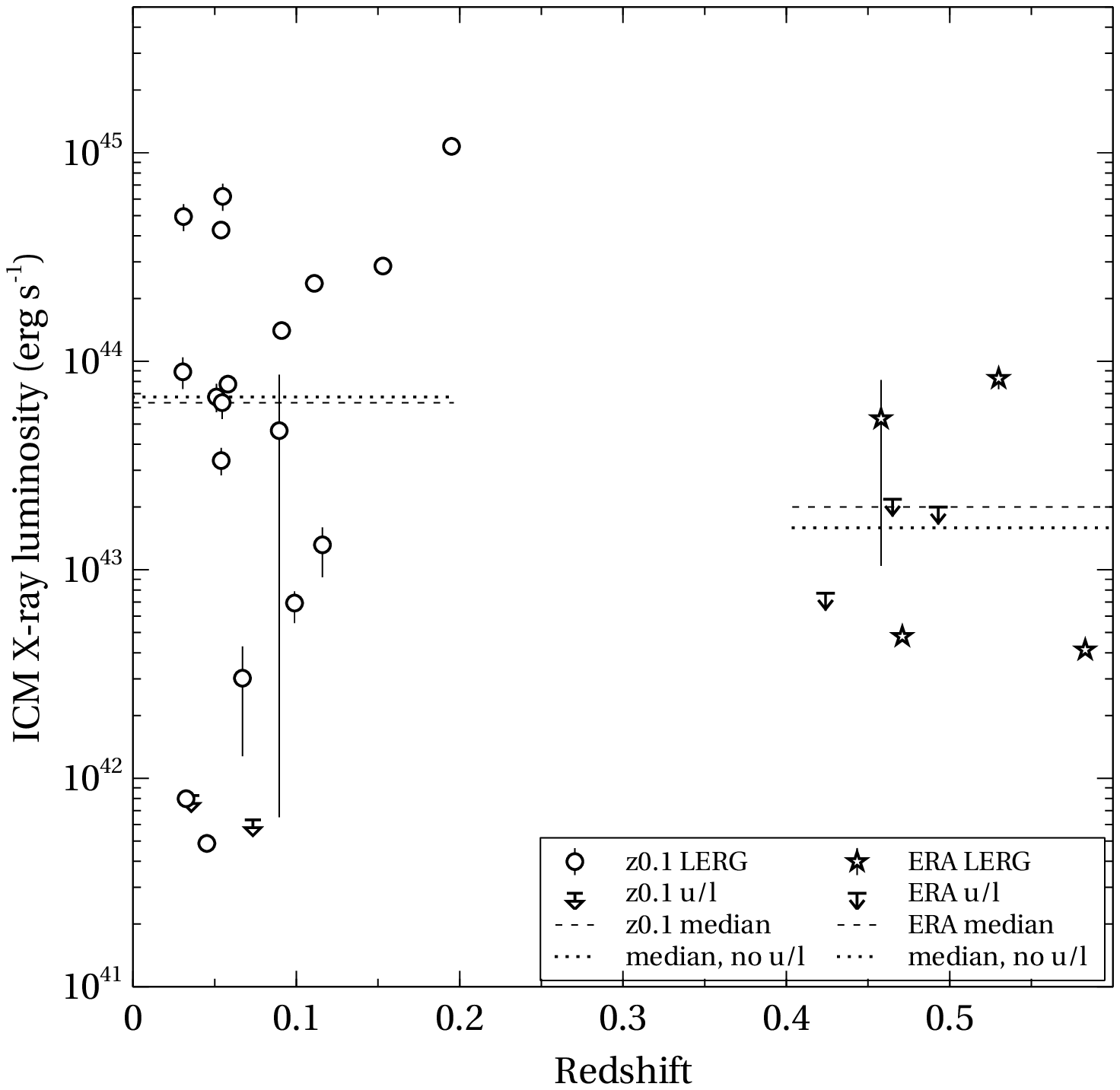}
  \end{minipage}
  \begin{minipage}{8cm} 
  \includegraphics[width=7cm]{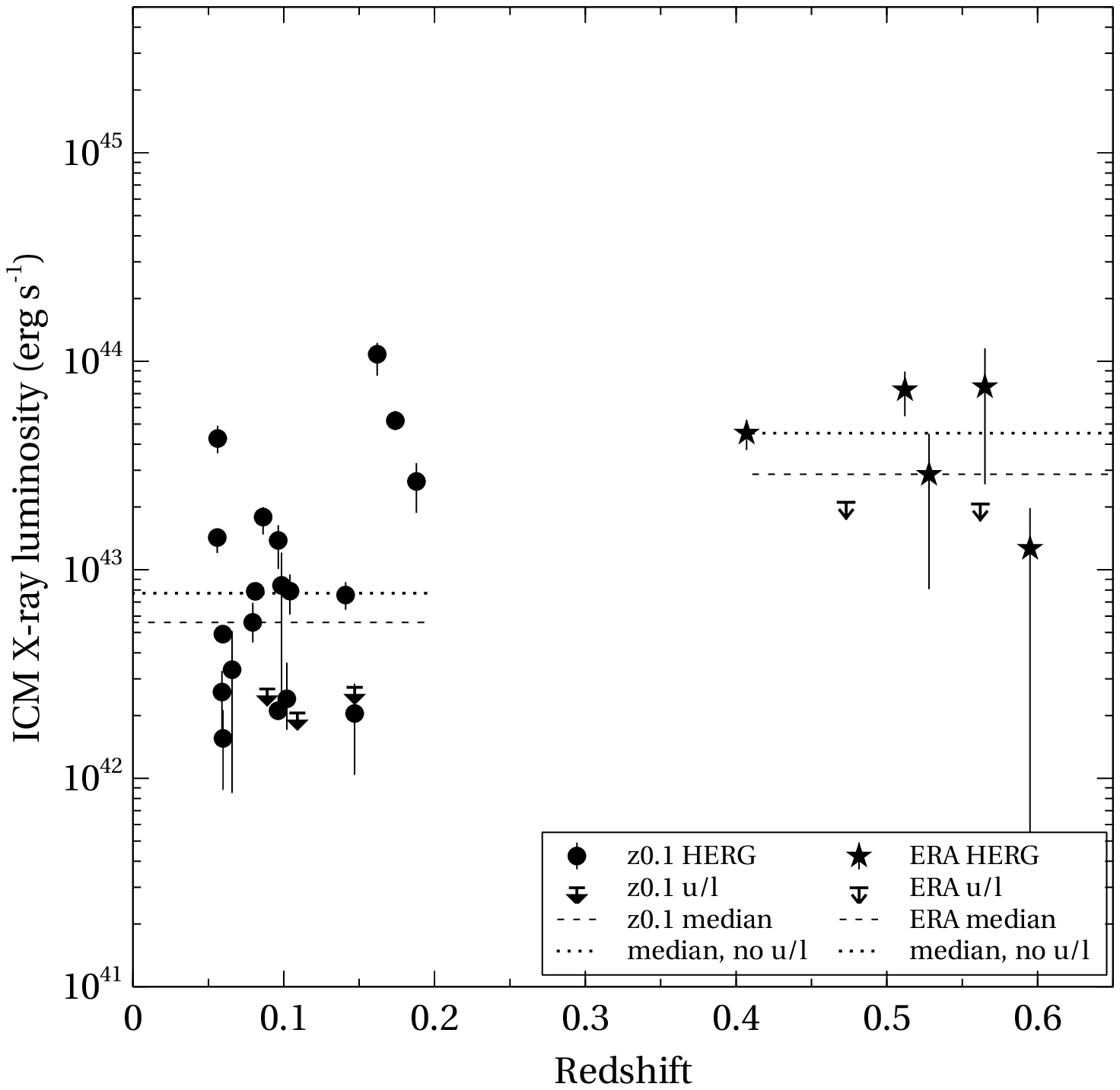}
  \end{minipage}
  \caption{ICM X-ray luminosity vs redshift for sub-samples matched in radio luminosity of the z0.1 and ERA samples, with LERGs on the left and HERGs on the right. Symbols as in Figure~\ref{fig:LrLxAll}.}
\label{fig:zLxM}\end{figure*}

We used Peto \& Prentice generalised Wilcoxon tests \citep{fei85} to look for differences in the median values of luminosity of the ERA and z0.1 LERG and HERG sub-samples (Table~\ref{tab:PandP}; the medians are shown in Figure~\ref{fig:zLx}). The tests showed no difference between the two LERG sub-samples, but a strong difference between the HERG sub-samples ($>3\sigma$). We repeated the tests with the sub-samples matched in radio luminosity (Figure~\ref{fig:zLxM}), and found the same but weaker trends.

\begin{table}
\caption{Tests for differences in sample median ICM luminosities, using the Peto-Prentice two-sample test}
\begin{tabular}{llcccc}\hline
Sample&Sub-sample&N1&N2&statistic&p\\
\hline
Full&All&55&26&1.8760&0.0607\\
&HERG&25&15&3.6860&0.0002\\
&LERG&30&11&0.1640&0.8696\\
Matched&HERG&22&7&2.2800&0.0226\\
&LERG&20&7&1.5340&0.1249\\
\hline\end{tabular}

N1 and N2 are the sizes of the samples being compared; p is probability under the null hypothesis.
\label{tab:PandP}\end{table}

We therefore have no evidence of evolution of the LERG population since $z\sim0.5$, but for the HERGs, although the maximum environment richness of the HERG population in the matched samples is the same at both redshifts, there are HERGs in poorer environments at $z\sim0.1$ than at $z\sim0.5$. Is this effect genuine, or is it due to non-detection of weak environments at high redshift?

We would need observations of almost 1000~ks to measure ICM luminosities of $10^{42}$~erg~s$^{-1}$ at the redshifts of the ERA sample, so we cannot know whether the lack of high redshift sources in this region of the plot is real or due to insufficient observation time. We have a hint that there might be at least occasional objects in this region from \citet{alm13} -- they obtained an extremely low galaxy-quasar spatial covariance function ($B_{gq}$) for PKS1136$-$13, which has a radio luminosity of $2.7\times10^{26}$~W~Hz$^{-1}$~sr$^{-1}$. If this source follows the expected correlation between $L_X$ and $B_{gq}$, it would then lie well into the lower right portion of the $L_X-L_R$ plot. The 80~ks X-ray observation of this source shows no sign of ICM emission, but this observation is not long enough to detect emission below $10^{43}$~erg~s$^{-1}$. However, this is the only 2Jy source in the high redshift range with such a low $B_{gq}$ --- the others have $B_{gq}$ values that should place them in environments around and above $10^{43}$~erg~s$^{-1}$. We also have only three upper limits in the ERA sample HERGs, and two of these are from short observations, so if there are weak environment sources at these redshifts then they are rare and are unlikely to have a dramatic affect on the sample medians. We therefore conclude that evolution of the HERG environment is probable but not certain.

\subsubsection{Cluster morphology}

The distributions of $\beta$ and core radius are shown in Figure~\ref{fig:betas}. The HERG and LERG sub-samples have slightly different $\beta$ medians -- 0.42 vs 0.52 -- but a Wilcoxon-Mann-Whitney test shows no significant difference.

The median core radius for the full z0.1 sample is also different for the two sub-samples -- 17 kpc for the HERGs and 62 kpc for the LERGs. In this case, the distributions of the HERGs and LERGs are different, and this is confirmed by the Wilcoxon-Mann-Whitney test (z=2.41, p=0.008), suggesting that the HERGs may have a higher concentration of gas near the cluster centre. However, when the core radius was scaled by $R_{500}$ the difference is no longer very significant (z=1.79, p=0.037), so the difference in core radius may be due to the difference in mass distributions between HERGs and LERGs rather than cluster shape.

\subsubsection{Central density}

If, as discussed in the Section~\ref{sec:intro}, LERGs follow a cycle fuelled by gas from the ICM and controlled by the central entropy of the system, we would expect the jet power to be related to the central conditions. We would also expect that, since ICM luminosity is related to cluster mass, $n_e$ should be related to the ICM luminosity. As discussed in Section~\ref{sec:spat}, we used the electron density $n_e$ at $0.1R_{500}$ as an indicator of the central conditions and compared it with radio luminosity.

\begin{figure*}
  \begin{minipage}{8cm}
  \includegraphics[width=7cm]{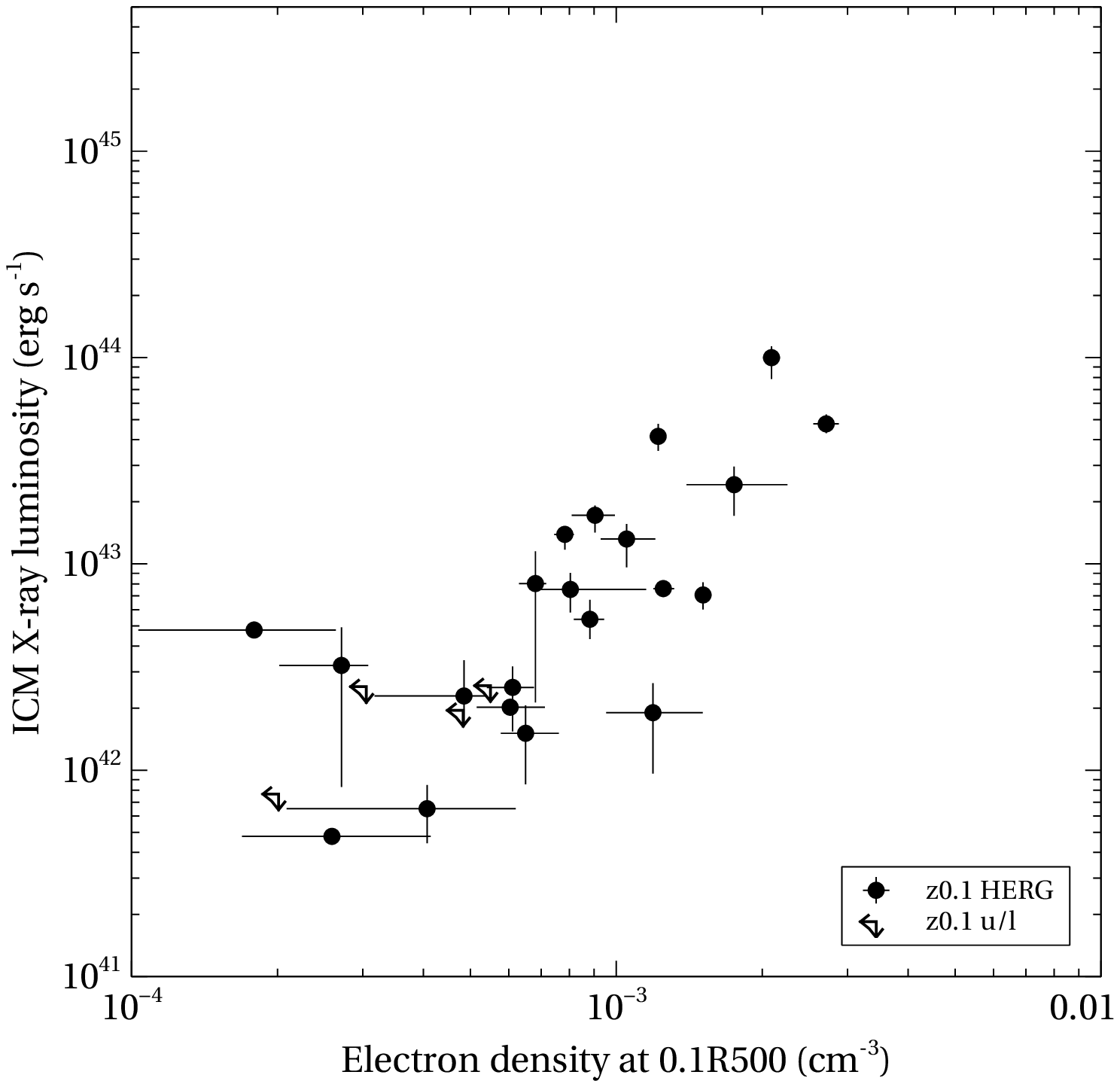}
  \end{minipage}
  \begin{minipage}{8cm} 
  \includegraphics[width=7cm]{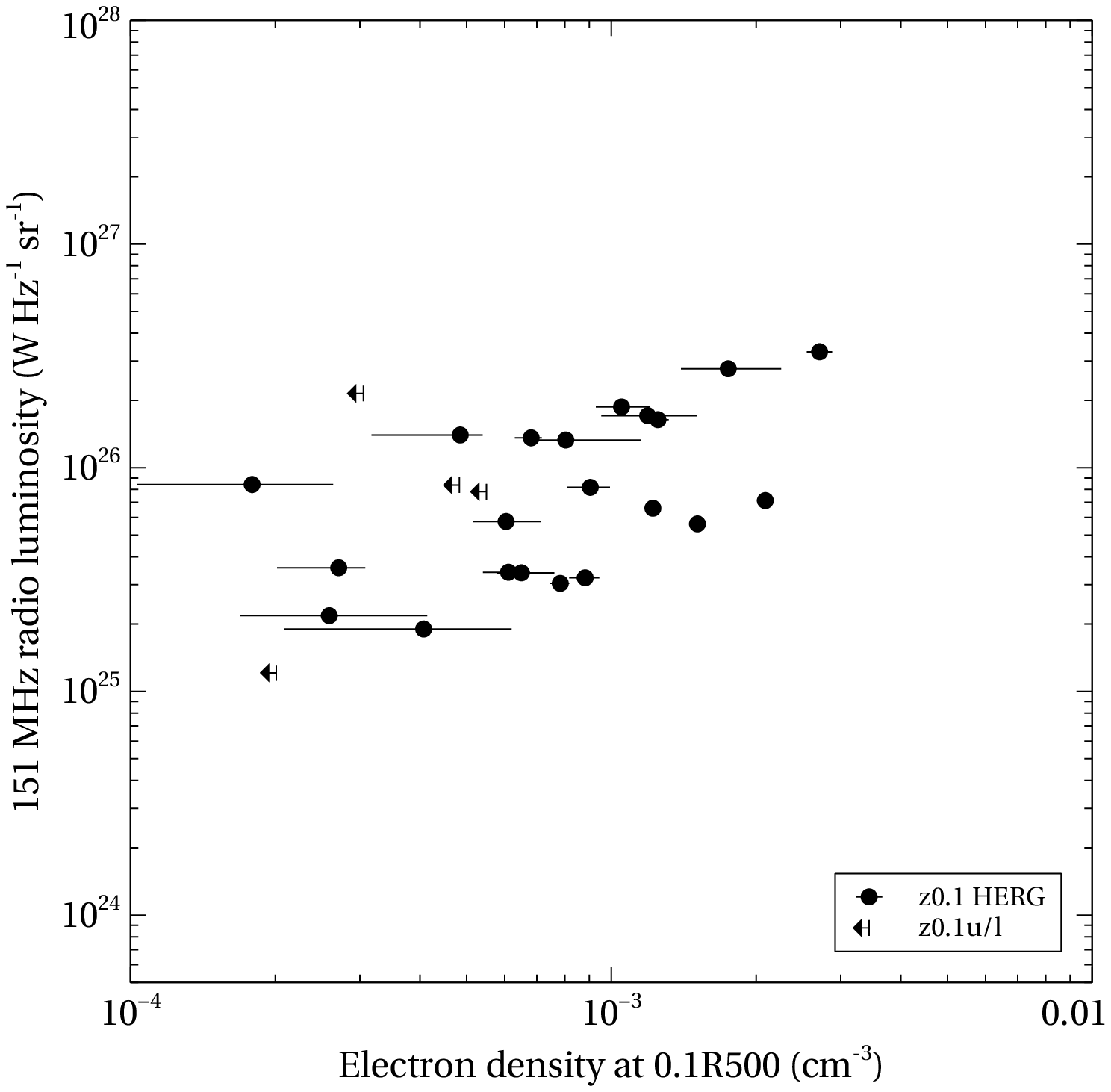}
  \end{minipage}
  \caption{Density at $0.1R_500$ vs ICM X-ray luminosity(left) and radio luminosity (right) for HERGs in the z0.1 sample. Symbols as in Figure~\ref{fig:LrLxAll}.}
\label{fig:neH}\end{figure*}

\begin{figure*}
  \begin{minipage}{8cm}
  \includegraphics[width=7cm]{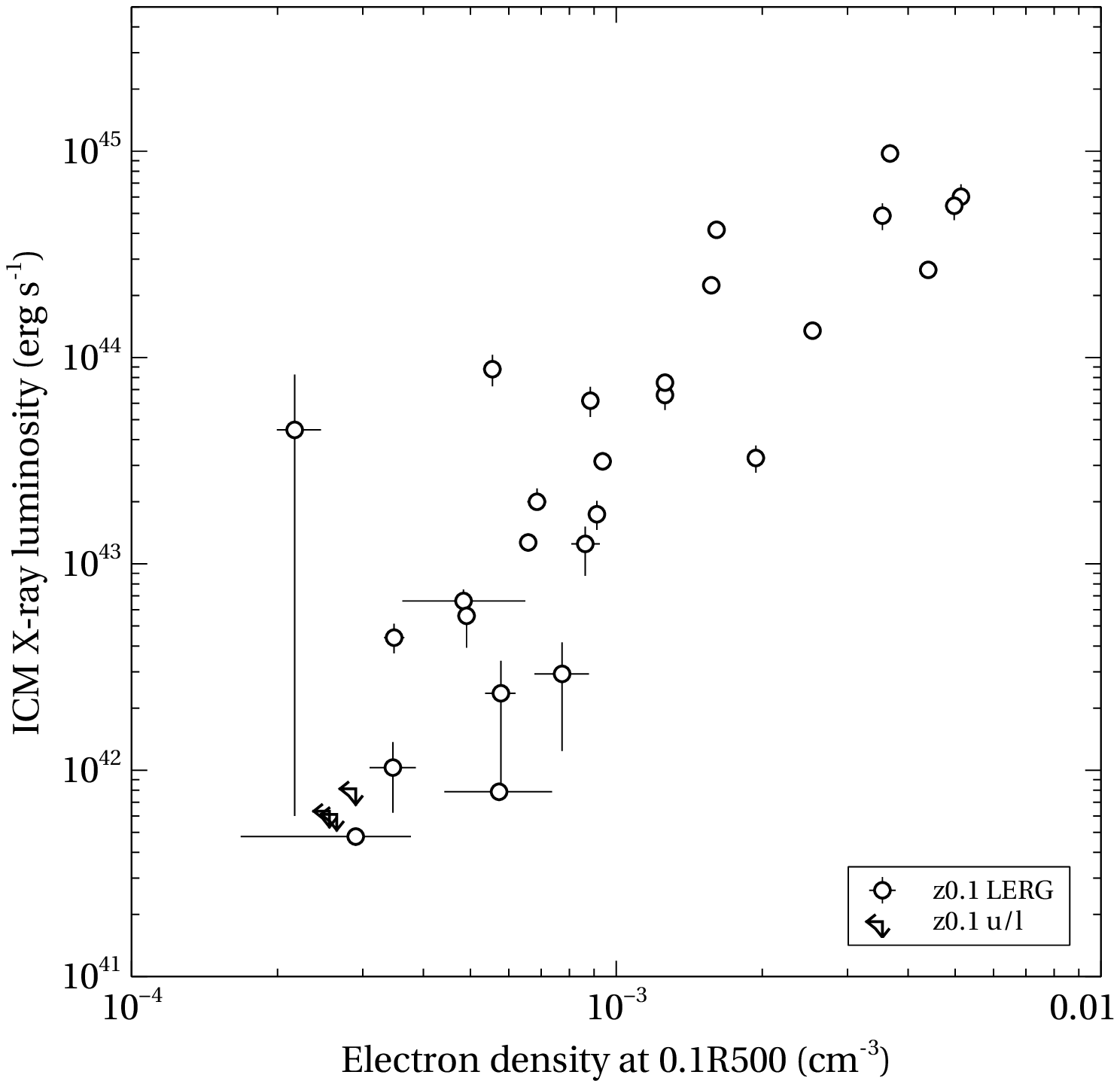}
  \end{minipage}
  \begin{minipage}{8cm} 
  \includegraphics[width=7cm]{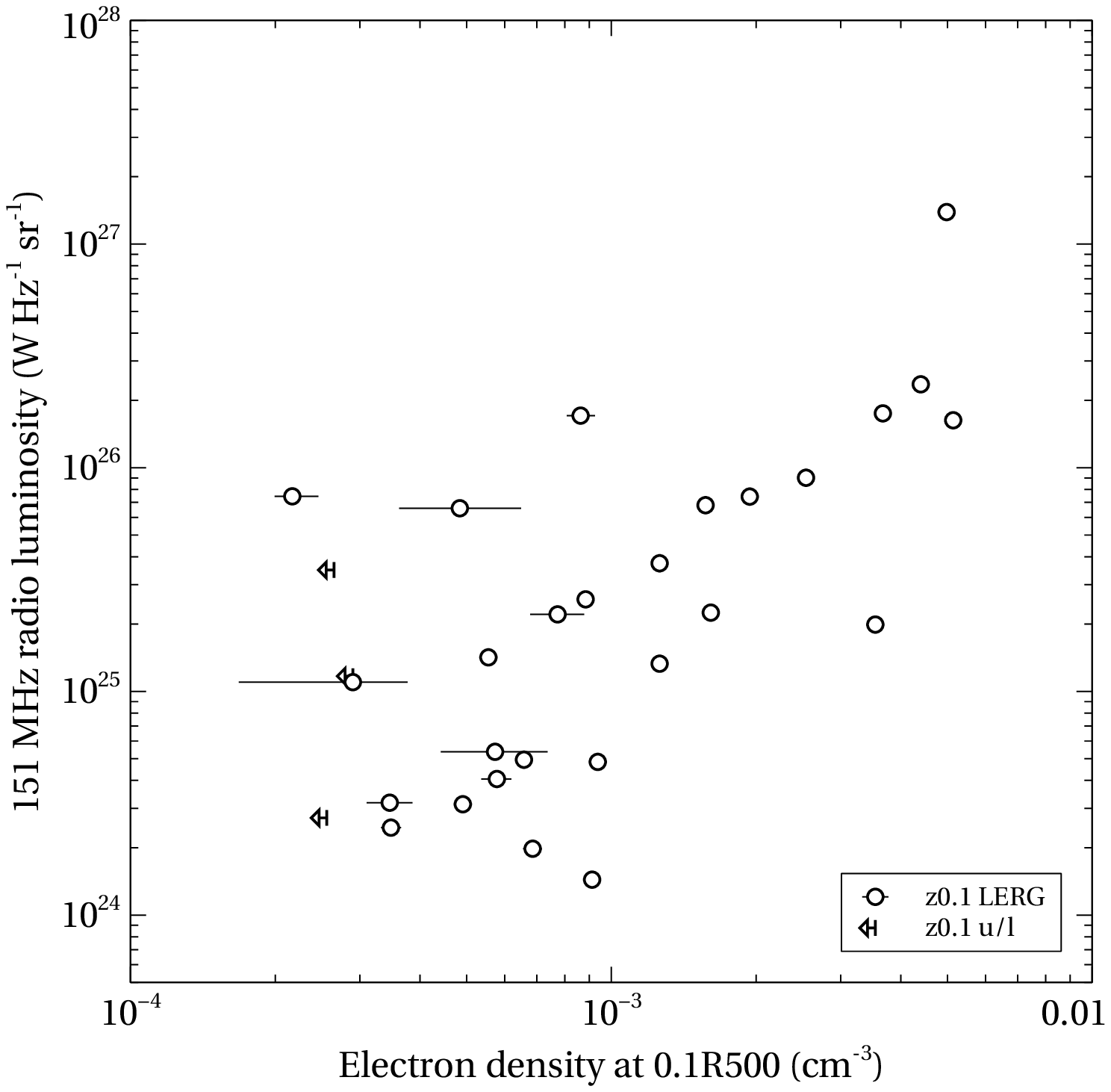}
  \end{minipage}
  \caption{Density at $0.1R_500$ vs ICM X-ray luminosity(left) and radio luminosity (right) for LERGs in the z0.1 sample. Symbols as in Figure~\ref{fig:LrLxAll}.}
\label{fig:neL}\end{figure*}

Figure~\ref{fig:neH} shows the electron densities at $0.1R_{500}$ plotted against ICM luminosity and radio luminosity for the HERGs, and Figure~\ref{fig:neL} for the LERGs for the z0.1 sample (we did not include the ERA sample as at their redshift the angular size of $0.1R_{500}$ was too close to the PSF to obtain reliable densities). As expected, the central density correlates strongly with ICM luminosity for both types of radio galaxy. In addition, there appears to be a relationship between $L_R$ and $n_e$ for the LERGs, but not for the HERGs, and this is confirmed at the 99.5~per~cent confidence level by the Generalised Kendall's Tau tests comparing $L_X$ and $L_R$ with $n_e$ in the presence of a common dependence on redshift (Table~\ref{tab:KTau}).

This result could simply be a reflection of the $L_R-L_X$ relationships for the two galaxy types, but if jet power is related to central density for the LERGs, this would also contribute to the $L_R-L_X$ relation. We therefore looked for a correlation between $L_R$ and $L_X$ in the presence of a common dependence on $n_e$. If the jet power is in a large part controlled by central density, this should remove the correlation; if jet power and central density are unrelated then the correlation should be unchanged. As can be seen from Table~\ref{tab:KTau}, the partial correlation test of $L_R-L_X$ in the presence of $n_e$ shows no significant correlation. However, performing the inverse test of looking for a correlation between $L_R$ and $n_e$ in the presence of a common dependence on $L_X$ also removed the correlation, suggesting that all three factors were well correlated.

In an attempt to find which of the three possible relationships ($L_R-L_X$, $L_R-n_e$ and $L_X-n_e$) was dominant, we did a Principal Components Analysis (PCA, \citealt{fra99}) on the data with the upper limits excluded, having checked that this made little difference to the results of the Generalised Kendall's Tau tests. As can be seen from Table~\ref{tab:PCA}, the three factors all contribute in similar proportions to the main principal component PC1, suggesting that we cannot determine the dominant relationship from this data. This was confirmed using Spearman's $\rho$ tests comparing the results of the full PCA with similar analyses on pairs of factors, which all gave similar and very strong correlations.

\begin{table}
\caption{Principal Component Analysis for radio luminosity, ICM luminosity and central density, using the $z0.1$ sample.}
\begin{tabular}{lccc}\hline
&PC1&PC2&PC3\\
\hline
Eigenvalue&1.924&0.487&0.205\\
Proportion&0.736&0.186&0.078\\
Cumulative&0.736&0.922&1.000\\
\\
Variable&PC1&PC2&PC3\\
\hline
Radio luminosity&0.495&0.646&0.581\\
ICM luminosity&0.646&0.296&0.407\\
Central density&0.581&-0.703&0.705\\

\hline\end{tabular}

PC1, PC2 and PC3 are the principal components in order of contribution to the total variance.
\label{tab:PCA}\end{table}

\subsection{Comparison with general cluster and group environments}

We assume in our analysis that the cluster environments of our radio-loud AGN do not differ markedly from those of clusters of similar luminosity without radio galaxies. We therefore compared our temperature-luminosity scaling relation with $L_X-T_X$ relations for other samples.

\begin{figure}
  \includegraphics[width=7cm]{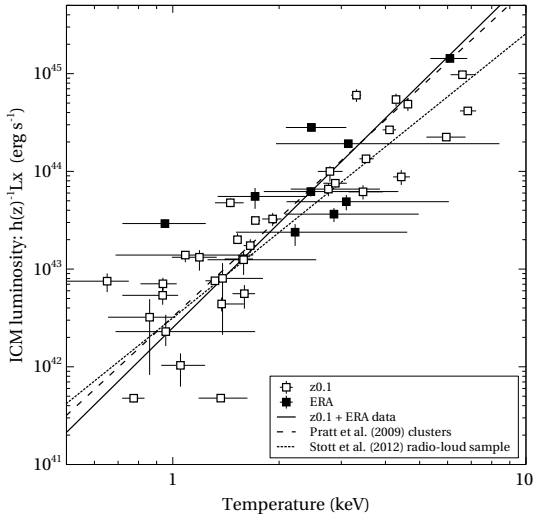}
  \caption{ICM luminosity vs temperature for the temperatures obtained by spectral analysis. Empty squares denote the z0.1 sample and filled squares the ERA sample. The solid line shows the $L_X-T_X$ relationship from the combined samples, and the dashed line shows \citet{pra09}'s $L_X-T_X$ relation for clusters. Also shown is the relation from \citet{sto12} for their sub-sample of clusters containing radio galaxies (dotted line).}
\label{fig:LxTx}\end{figure}

There is a very strong correlation between ICM temperature and luminosity (Table~\ref{tab:KTau}). Figure~\ref{fig:LxTx} shows ICM temperature plotted against luminosity for all sources in the z0.1 and ERA samples with temperatures obtained by spectral analysis. We used the orthogonal BCES method from \citet{akb96} to calculate the regression line (solid line, $log_{10}L_X=(3.56\pm{0.36})log_{10}T_X+(42.40\pm{0.15})$).

The dashed line shows the \citet{pra09} $L_X-T_X$ scaling relation for the REXCESS clusters, which we used to obtain the estimated temperatures. Our slope is slightly higher ($3.56\pm{0.36}$ vs $3.35\pm{0.32}$), but compatible. Also plotted in Figure~\ref{fig:LxTx} is the $L_X-T_X$ relation obtained be \citet{sto12} for clusters containing radio sources from the \textit{XMM} Cluster Survey. Their slope of $2.91\pm{0.45}$ is lower than our result.

Our sample contains both galaxy groups and clusters, both cool-core and non-cool-core environments and consists entirely of radio galaxies. These factors have all been found to modify the $L_X-T_X$ relation (e.g. \citealt{hel00,cro05,che07,mag07,pra09,eck11,sto12,bha15}). In addition, selection biases (in particular the Malmquist bias, which ensures that for a given temperature, the most luminous objects are preferentially selected) can have a strong effect on the slope of the $L_X-T_X$ relation (e.g. \citealt{eck11,bha15}), and the evolution parameter, which is usually assumed to be unity, is likely to be affected by the break in self-similarity in the $L_X-T_X$ relation \citep{mau14}.

Our relation is likely to be affected by all the factors mentioned above, but lies near or within most of the ranges cited by the literature and so our sample shows no evidence of being different from other cluster and group samples.

\begin{figure}
  \includegraphics[width=7cm]{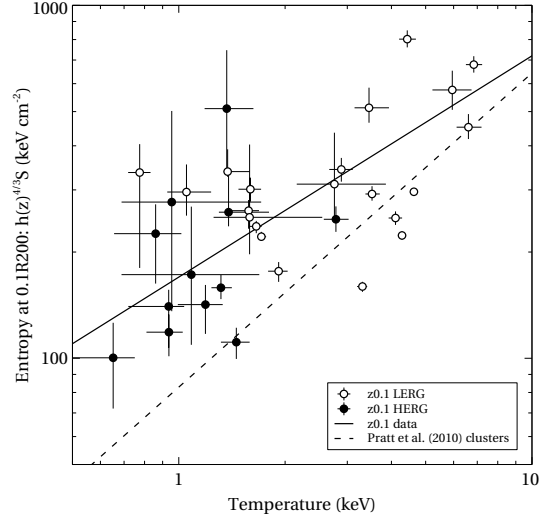}
  \caption{Entropy vs temperature for the z0.1 sample temperatures obtained by spectral analysis. Empty circles are LERGs and filled circles are HERGs. The solid line shows the $S-T_X$ relationship from the z0.1 sample and the dashed line shows \citet{pra10}'s $S-T_X$ relation for clusters.}
\label{fig:STx}\end{figure}

We also calculated entropy $S$ within $0.1R_{200}$ for the sources from the z0.1 sample with temperatures obtained by spectral analysis. We used $h^{4/3}(z)S=kT/{n_e^{2/3}}$, where $R_{200}$ is the radius at an overdensity of 200 \citep{arn05}, $kT$ is the ICM temperature and $n_e$ is the electron density. $n_e$ was calculated as described in Section~\ref{sec:spat}. The results are shown in Figure~\ref{fig:STx}.

We obtained a regression line for the $S-T_{X}$ relation, again using the \citet{akb96} orthogonal BCES method, and obtained a shallower slope than that of \citet{pra10} ($0.63\pm{0.13}$ vs $0.89\pm{0.15}$). Our slope does however lie within the range of results from the literature cited by \citet{pra10} (slopes of 0.49 to 0.92). If, as discussed in Section~\ref{sec:intro}, interactions form part of the triggering process for HERGs, we would expect them to have high entropy for their temperature. \citet{pra10} noted that disturbed clusters tend to have high entropy compared with relaxed clusters of the same temperature, and indeed all but one of our HERG sample lie above their regression line.

Overall, there is no systematic evidence that the luminosities of our sample of radio galaxies differ from those of galaxy groups and clusters that do not host radio-loud AGN, so our use of luminosity as a proxy for total cluster mass is reasonable. Our entropies tend to be high, which is likely to be at least in part due to clusters being disturbed, but are still within the ranges cited in the literature. 

\subsection{Comparison with optical measures}\label{sec:opt}

\begin{figure}
  \includegraphics[width=7cm]{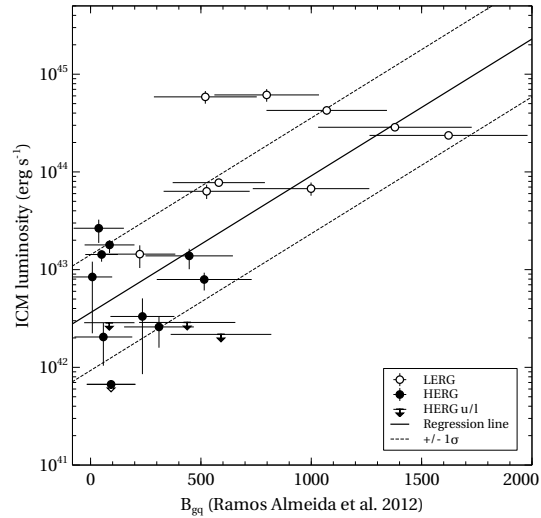}
  \caption{ICM luminosity vs galaxy-quasar spatial covariance function $B_{gq}$ for the 2Jy sources in the z0.1 sample. $B_{gq}$ values were taken from \citet{alm13}. Empty circles are LERGs and filled circles are HERGs. The solid line shows the regression line for the data, and the dotted lines the $1\sigma$ scatter.}
\label{fig:LxBgg}\end{figure}

We would expect a correlation between different measures of cluster richness, though with a fair amount of scatter (e.g. \citealt{yee03,led03}, I13). \citet{alm13} have calculated the galaxy-quasar spatial covariance function ($B_{gq}$) for the 2Jy sample. We therefore compared these with the ICM luminosities for the 2Jy sources within our z0.1 sample (Figure~\ref{fig:LxBgg}).  A generalised partial $\tau$ showed a correlation at the $3\sigma$ level (Table~\ref{tab:KTau}). We used the Buckley-James method \citep{iso86} to obtain a regression line including the upper limits in $L_X$; this gave $log_{10}L_X=(0.0014\pm{0.0003})B_{gq}+(42.58\pm{0.59})$. The amount of scatter compromises the relationship's utility as a scaling relation, and the two measures were taken within different radii (170~kpc for $B_{gq}$ and $R_{500}$ (median 600~kpc) for $L_X$). However, the strength of the correlation shows that overall results from the two measures should be comparable. 

\subsection{Implications}

Our results add evidence for a difference in cluster environments to the increasing wealth of data supporting a dichotomy in the population of radio-loud AGN, and provide supporting evidence for models that involve the different accretion cycles of the high and low excitation sources. This difference in cluster environments may also play a part in modifying radio galaxy polarisations, as suggested by \citet{osu15}, who recently reported a dichotomy in the integrated degree of polarisation of HERGs and LERGs.

\subsubsection{LERG results}

The strong correlations between radio luminosity, ICM luminosity and central density for LERGs supports the concept of some form of relationship between ICM properties and jet power. This could provide corroborative evidence for a stable, long-lasting feedback cycle as demonstrated in the simulations of \citet{gas12,gas13}, in which matter is driven outwards by the jets and falls inwards from the ICM when the central conditions become favourable. The pressure of available gas in the ICM will affect the flow of gas into the nucleus, which will in turn affect the accretion rate and the jet power. There will be some variation as the central regions heat up and then cool again as the gas clumps accrete and disperse and are recharged, and this may be one of the several factors that could contribute to the scatter in the $L_{R}-L{_X}$ plot. Other possible factors that might add scatter to the relationship are discussed briefly in I13 -- the scatter in the relationship between mechanical jet power and total radio luminosity (e.g. \citealt{bir08,cav10,god13,hak13}) and the effect of differences in central gas properties (which are not correlated with ICM luminosity, e.g. \citealt{cro08b}) on jet power. It might be that the cyclic variations suggested by \citet{gas12,gas13} could contribute to both these issues.

An alternative explanation for the correlation is that it results at least in part from `environmental boosting' (e.g. \citealt{bar96,hak13}), where for a given jet power, a richer environment increases the radio luminosity. However, this effect should apply equally to both HERGs and LERGs. The fact that the individual HERG samples show no correlation suggests that an explanation based on fuelling mechanism such as that described above is more likely. It also suggests that the environmental boosting may be counteracted by another mechanism. Entrainment is expected to increase with environment density and reduces radio luminosity relative to jet power (eg. \citealt{cro14}), and so works in the opposite direction to environmental boosting.

\subsubsection{HERG results}

Turning to the HERGs, the lack of correlation between the radio luminosity and cluster properties for the individual samples suggests that the ICM does not have a major r\^ole to play either in powering the system or in controlling the luminosity of the lobes. This fits in with the theory that HERGs accrete at a relatively high rate from an accretion disc maintained by a local reservoir of cold gas, probably originating from galaxy interactions and mergers (e.g. \citealt{har07b,alm12,tad14}). Gas ingestion from the ICM as described by \citet{gas13} is sufficient to fuel even the most powerful LERG jets, and is also sufficient to fuel low-power HERGs. It could not however maintain the accretion rates required to power the jets and nucleus of more powerful HERGs. The richness of the ICM would therefore be largely irrelevant and no correlation is to be expected.

If the difference between the HERG environments in the ERA and z0.1 environments does indicate evolution, then this is very interesting. At high redshifts ($z>1$), radio galaxies are typically seen in richer environments than similar radio-quiet galaxies \citep{wyz13,hat14}, but by $z\sim0.5$ we are finding radio galaxies in groups and poor clusters. This may be due to cosmic downsizing --- as the gas density in the vicinity of the black hole depletes as a result of AGN activity and star formation, then the conditions required for HERGs to be triggered, such as mergers bringing new gas into the host galaxy, occur in poorer environments as redshift reduces. However, although the minimum environment richness at $z\sim0.1$ is lower than at $z\sim0.5$ (as would be expected from cosmic downsizing) the maximum richness of the environments is the same. Thus there must be additional processes involved.

\section{Conclusions}

We have compared low frequency radio luminosity with the richness of the cluster environment for a sample of 55 radio-loud AGN lying in the redshift range $0.01\leq z\leq 0.2$. The sample covered three decades of radio luminosity and contained 25 high excitation and 30 low excitation sources. We used the X-ray luminosity of the ICM as the measure of cluster richness. The X-ray observations were taken from the {\it Chandra} and {\it XMM-Newton} archives.

We found:
\begin{itemize}
\item For the complete z0.1 sample, a weak correlation between radio luminosity and cluster richness;
\item For the LERG sub-sample, strong correlations between radio luminosity, cluster richness and central density;
\item For the HERG sub-sample, no correlation between radio luminosity and cluster richness, or between radio luminosity and central density;
\item The core radii of the LERGs were on average larger than those of the HERGs.
\end{itemize}

These results were similar to those of our previous study (I13), which used the ERA sample, containing 26 sources at higher redshift ($0.4 \leq z \leq 0.6$). We compared the results for $z\sim0.1$ and $z\sim0.5$, and found:

\begin{itemize}
\item For the two LERG samples, the slopes and normalisations of the radio luminosity--ICM richness correlation were very similar, giving a very strong correlation. This suggested that there had been no evolution with redshift since $z\sim0.5$, and we found no significant difference between the cluster environments of the two samples;
\item Although the HERGs in both samples had similar maximum environment richnesses, the lower redshift sample also included environments much weaker than any seen in the ERA sample, giving tentative evidence of evolution of the environment;
\item The HERGs occupied groups and clusters with more compact central regions than the LERGs, suggesting that HERGs may have a greater concentration of gas near the host galaxy than LERGs;
\item We found the luminosity-temperature relation for our samples to be compatible with $L_X-T_X$ relations derived for general samples of clusters and groups, suggesting that the environments occupied by radio-loud AGN are similar to those of typical clusters.
\end{itemize}

The evidence of a difference between the HERG and LERG large-scale environments is now strong, so we next intend to investigate relationships between the ICM and other factors where differences have been found between the two types of AGN.

\section*{Acknowledgments}
JI acknowledges the support of the South-East Physics Network (SEPnet). JHC acknowledges support from the Science and Technology Facilities Council (STFC) under grant ST/J001600/1.

The scientific results reported in this article are based on observations made with the {\it Chandra} X-ray observatory and on observations obtained with {\it XMM-Newton}, an ESA science mission with instruments and contributions directly funded by ESA Member States and NASA. This research has made use of software provided by the {\it Chandra} X-ray Center (CXC) in the application packages \textsc{ciao} and \textsc{chips}, and of the {\it XMM-Newton} Science Analysis Software (\textsc{sas}). This research also made use of the University of Hertfordshire high-performance computing facility (http://stri-cluster.herts.ac.uk/). Figures were plotted using the \textit{veusz} package (http://home.gna.org/veusz/).

\appendix
\section[]{Notes on individual sources}

\subsection{3CRR sources}

\textbf{3C~28} (A115-N, PKS~0053+26) is a LERG with FRII morphology. Its host is one of a pair of merging clusters and is highly asymmetric, and the disturbance is visible in the surface brightness profile. Both clusters are rich -- \citet{gut05} obtained values of around 5~keV for the individual clusters away from the merger and 8~keV for the plasma between the clusters. Because we are interested in the full environment around the host galaxy, we included parts of both regions and obtained a temperature of $6.61^{+0.59}_{-0.49}$~keV. This lies within the upper bound of the temperature of $5.35^{+1.28}_{-0.09}$~keV obtained by \citet{she11}.

Our luminosity lies very close to the value expected from the $L_X-T_X$ relation. It is higher than that obtained by \citet{she11} --- our surface brightness profile is substantially wider, perhaps because of our use of double background subtraction and double~$\beta$ modelling.\\

\noindent
\textbf{3C~31} (NGC~383) is a LERG with a massive central rotating disc of molecular gas \citep{oku05} and FRI morphology. Our temperature of $1.53\pm{0.3}$~keV is consistent with the those obtained by \citet{kom99} and \citet{cro14}. \citet{hod10} found a higher temperature ($2.0^{+0.5}_{-0.02}$~keV) in the central regions with a \textit{Chandra} observation.\\

\noindent
\textbf{3C~33} (PKS~0106+13) is a Narrow Line Radio Galaxy (NLRG) with FRII morphology lying in a weak environment. The radio outburst is thought to be energetic enough to eject a significant fraction of the corona gas \citep{kra07}. We detected very little emission beyond the host galaxy. The counts are low so we obtained an estimated temperature (1.1~keV). This is typical of a group environment.\\

\noindent
\textbf{3C~35} is also an NLRG with FRII radio structure in a weak environment. \citet{man13} report a gas `belt' lying around the source between the lobes and extending out to about 170~kpc -- not far short of our detected radius of emission. We found a broad surface brightness profile and an estimated temperature of 1~keV, very similar to the spectral temperature obtained by \citet{man13} using combined \textit{Chandra} and \textit{XMM-Newton} observations.\\

\noindent
\textbf{3C~66B} is a LERG with FRI morphology. Our temperature of 1.7~keV is compatible with that of \citet{cro08a}, but we obtained a higher luminosity ($3.17^{+0.10}_{-0.12}$ vs $1.07^{+0.08}_{-0.24}\times{10}^{43}$~erg~s$^{-1}$). The reason for the discrepancy is not clear, but could relate to the use of an improved background subtraction method in this work.\\

\noindent
\textbf{3C~76.1} is a LERG with FRI plumes spreading across a large proportion of the detected environment, which we found to be weak. \citet{mil99}, using the \textit{ROSAT} All-Sky Survey, did not detect any extended emission. Because of the low counts (this was a short observation), we could not fit a double-$\beta$ model. We used an estimated temperature of 0.65 keV. This is lower than the $0.91^{+0.25}_{-0.14}$~keV obtained by \citet{cro08a} with an \textit{XMM-Newton} observation; our luminosities are however compatible.\\

\noindent
\textbf{3C~98} (PKS~0356+10) is an NLRG with FRII morphology in a weak environment with the detected emission not extending far beyond the host galaxy. Again, \citet{mil99} did not detect extended emission. We used an estimated temperature of 0.62~keV. This was lower than the $1.1^{+0.3}_{-0.2}$~keV obtained by \citet{hod10} within a smaller radius.\\

\noindent
\textbf{3C~192} (PKS~0802+24) is another NLRG with FRII morphology in a weak environment. This was a short observation, and although the emission extended beyond the host galaxy, there were insufficient counts to fit a double-$\beta$ model. We used an estimated temperature (0.8~keV).\\

\noindent
\textbf{3C~219} is a Broad Line Radio Galaxy (BLRG) with FRII morphology. It may be recently re-triggered after a dormant period \citep{wei14} as it has a small jet within a well-formed pair of lobes with a classical double structure.

The temperature (1.5~keV) and luminosity ($5.2\times{10}^{43}$~erg~s$^{-1}$) are those of a group, and are compatible with those obtained by \citet{she11}.\\

\noindent
\textbf{3C~236} is a LERG with massive FRII lobes extending well beyond the imaging chips. It is a double-double, with outer lobes of 4~Mpc in extent and inner lobes of 2~kpc \citep{tre10}. It is thought to have been reactivated about $10^5$~years ago \citep{lab13}. We found a weak environment, with insufficient counts for spectral analysis. Our estimated temperature of 1.2~keV is that of a group environment.\\

\noindent
\textbf{3C~285} is an NLRG with FRII morphology in a highly disturbed group environment. The host galaxy is also highly disturbed, probably as a result of a recent merger \citep{all02}, and is currently interacting with another galaxy \citep{bau98}. We obtained a temperature of $0.94^{+0.10}_{-0.22}$~keV for the extended emission, consistent with \citet{hod10}. \citet{har07c} obtained a lower temperature of 0.64~keV with a $\textsc{mekal}$ model; we found compatible results when we replaced our $\textsc{apec}$ model with a $\textsc{mekal}$ model.

Our surface brightness profile, generated with a double $\beta$ model and double background subtraction, gave a wider profile than that obtained by \citet{har07c}, and our luminosity is consequently higher -- $0.56^{+0.14}_{-0.11}$ vs $0.19^{+0.1}_{-0.1}\times{10}^{43}$~erg~s$^{-1}$.\\

\noindent
\textbf{3C~293} is a LERG with FRI plumes in a weak environment. It is a double-double source with estimated ages of $\sim20$~Myr for the outer lobes and $\la0.1$~Myr for the central source \citep{jos11}, and has strong, jet-driven outflows of gas \citep{mah13,lan15}. We obtained a temperature of 0.78~keV -- a weak group temperature. We found the emission to be under-luminous for its temperature. \\

\noindent
\textbf{3C~296} (NGC~5532, PKS~1414+11) is a LERG with FRI morphology and occupies a group environment -- \citet{mil99} found 4 galaxies in the group. The source is relatively near and the emission extends to just beyond the imaging chip. We therefore could not use double subtraction for modelling the profile, but were able to use one of the outer chips to model the background and so used double subtraction for the spectral analysis.

\citet{cro08a}, using an \textit{XMM-Newton} observation, obtained a temperature of 0.9~keV over a 50-600~arcsec region (larger than was available with our \textit{Chandra} observation), but found that the temperature was higher ($\sim 1.4$~keV), albeit with large errors, towards the centre. We also found that the temperature dropped gradually beyond around 150~arcsec, and so used the temperature within this region (1.6~keV). \citet{hod10} found the temperature to be 4~keV within a similar region, which is high compared with the results of \citet{cro08a} and our current work.\\

\noindent
\textbf{3C~303} is a BLRG with FRII structure at a steep angle to the viewer, and we found emission typical of a group environment. The observation had some pile-up, so we excluded the central 1.5~arcsec in the surface brightness profile. Our temperature and luminosity are slightly lower than those obtained by \citet{she11} -- $0.94^{+0.09}_{-0.13}$ vs $1.86^{+3.00}_{-0.54}$~keV and $0.76^{+0.12}_{-0.11}$ vs $0.92\pm{0.06}\times{10}^{43}$~erg~s$^{-1}$. Our luminosity is a little high compared with the $L_X-T_X$ relation, but within the scatter. \\

\noindent
\textbf{3C~305} is a NLRG with a halo around the host galaxy, probably of material being driven out by the jets \citep{har12}. It has unusual FRII lobes spreading at right-angles to the jets. In common with \citet{har12} and \citet{mil99}, we found little evidence of ICM emission beyond the host galaxy, and so we derived upper limits for this source.\\

\noindent
\textbf{3C~310} (PKS~1502+26) is a LERG with wide FRI plumes and a disturbed environment containing a large cavity, filamentary structure and a shock front at about 180~kpc from the nucleus \citep{kra12}. The emission extends beyond the observing chip, but we were able to use one of the outer chips to model the background for the spectral analysis. We obtained a temperature of 1.9~keV, typical of a strong group or weak cluster. \\
 
\noindent
\textbf{3C~321} (PKS~1529+24) is an NLRG with FRII morphology in the process of merging with a neighbouring galaxy \citep{eva08}. The environment is weak and highly disturbed; we found little emission beyond the host galaxy and had low counts, so we used an estimated temperature of 0.87~keV.\\

\noindent
\textbf{3C~326} (PKS~1550+20) is a LERG with large FRII lobes extending beyond the imaging chip. It shows evidence of molecular reservoirs thought to be formed by positive feedback \citep{nes11}. The host galaxy has a nearby companion. We found the environment to be wide and flat and fairly weak. We had insufficient counts for spectral analysis so used an estimated temperature of 1.9~keV.\\

\noindent
\textbf{3C~338} (NGC~6166) is a LERG with FRI morphology, lying in the cluster Abell~2199. The cluster is highly disturbed, and \citet{nul13} discuss a variety of features in the cluster plasma including a shock front at 100~arcsec and a large plume extending to about 50~arcsec probably resulting from a cluster merger. These are visible in our surface brightness profile.

The emission extends well beyond the imaging chip so double subtraction was not possible for either the profile or the spectral analysis. \citet{nul13} give a detailed temperature map of the centre of the cluster showing complex structure, and we obtained temperatures rising from a cooler centre in line with their results. Our overall temperature of $4.63\pm{0.08}$~keV is consistent with those reported by \citet{kas99} (using a \textit{ROSAT} observation) and \citet{hod10}. \\

\noindent
\textbf{3C~346} (4C~17.70, PKS~1641+17) is an NLRG with FRI morphology lying in a weak cluster. Our temperature of 2.8~keV is consistent with that of \citet{she11}. Our profile shows the same features as that of \citet{she11}, but has a much larger detected radius and a shallower outer slope, perhaps due to improved background modelling from the double subtraction. Our luminosity is therefore substantially higher (10.8 vs 0.65$\times{10}^{43}$~erg~s$^{-1}$) and lies close to the $L_X-T_X$ relationship.\\

\noindent
\textbf{3C~386} (PKS~1836+17) is a LERG with broad $H\alpha$ lines in the spectrum \citep{sim96} and FRI lobes. The observation had low counts after background subtraction so we were unable to fit a double-$\beta$ profile. There were insufficient counts left after subtraction of the \textit{Chandra} blank sky files to obtain a background for the spectral analysis so we used single subtraction using a background from the observing chip to obtain the temperature, which was that of a group (1~keV).\\

\noindent
\textbf{3C~388} is a LERG in a cluster with cavities around its FRII lobes and a sub-cluster to the east \citep{kra06}. The emission extends beyond the imaging chips, but we were able to use a side chip to model the background. We found a cluster temperature of 3.5~keV.

Our temperature and metallicity for the extended ICM agree with the results of Kraft et al., but our temperature is slightly higher than that obtained by \citet{hod10}; this is probably due their use of a different (frozen) metallicity.\\

\noindent
\textbf{3C390.3}. This BLRG has fast gas outflows from the central regions \citep{tom10}. It has FRII morphology and is in a strong environment. The ICM emission extends beyond the imaging chip, and the observation also suffers from considerable pileup. The temperature we obtained using single subtraction with the \textit{Chandra} blank sky files was low for such bright emission ($<$1~keV). We had very few counts left after the \textit{Chandra} blank sky subtraction and could not analyse a background spectrum, so it is possible that the blank sky files were not accurate for this region or that the effects of the pileup extended across a wide energy range. We therefore used an estimated temperature for this source.\\

\noindent
\textbf{3C~433} (4C~24.54, PKS~2121+24) is an NLRG with hybrid FRI/FRII morphology in a group environment. The northern lobe is very bent, perhaps due to interaction with the surrounding ICM \citep{hod10}. We obtained a temperature consistent with that of \citet{hod10}, but slightly lower than that of \citet{she11}. Our luminosity is higher than that of \citet{she11} -- $0.24^{+0.12}_{-0.07}$ vs $0.05\pm{0.01}\times{10}^{43}$~erg~s$^{-1}$ -- but lies on the $L_X-T_X$ relationship. As with 3C~28, our modelling methods resulted in a much wider profile than that obtained by \citet{she11}.\\

\noindent
\textbf{3C442A} (PKS~2212+13) is a LERG with FRI plumes, hosted by the interacting galaxy pair NGC~7236/7237 \citep{wor07}, with filaments from the interacting galaxies and a ridge structure between the plumes \citep{har07c}. \citet{wor07} found no evidence of an active jet, and suggest that the jets may have been stopped by an excess of central gas pressure resulting from the merger. The system lies in a weak cluster, and extends beyond the imaging chip. We could not obtain a background from the outer chips so used the \textit{Chandra} blank sky files for the background for the spectral analysis. We obtained a temperature consistent with that of \citet{har07c}, and a luminosity close to the $L_X-T_X$ relation.\\

\noindent
\textbf{3C~449} is a LERG with FRI plumes extending well beyond the imaging chip, and is relatively close, so the environment emission extends beyond the chip. It has a 600~pc dust disk which, unusually, lies nearly parallel to the jet \citep{tre06}. The surface brightness profile is wide with an unusually shallow slope, though well constrained by the MCMC modelling.

Our temperature of $1.66^{+0.06}_{-0.07}$~keV is higher than that obtained from a \textit{XMM-Newton} observation by \citet{cro08a} ($0.98\pm{0.02}$~keV), but was measured over a smaller region. \citet{hod10} obtained a temperature of $1.58\pm{0.06}$~keV over a smaller region still; their modelling included the contribution of the host galaxy, which we excluded.

Our luminosity of $1.75^{+0.28}_{-0.28}\times{10}^{43}$~erg~s$^{-1}$ is slightly higher than that obtained by \citet{cro08a} ($1.20^{+0.12}_{-0.10}\times{10}^{43}$~erg~s$^{-1}$), but because of our higher temperature, our luminosity is calculated within a larger radius. \\

\noindent
\textbf{3C~452} is an NLRG with FRII morphology. There is a lot of X-ray emission associated with the radio structure, but beyond this the environment is poor. Having excluded the emission associated with the lobes, \citet{she11} found a temperature of $1.18\pm{0.11}$~keV in the inner 160~arcsec, and $0.86^{+0.13}_{-0.05}$~keV beyond. We also found that the temperature dropped as the radius increased, but had a slightly higher temperature of $1.32^{+0.10}_{-0.08}$~keV in the central region. \\

\noindent
\textbf{3C~465} is a LERG with large, bent FRI plumes lying in a cluster. The cluster has a cool core with the temperature increasing to around 5 keV before levelling off \citep{har05}. We obtained an overall temperature of $4.43^{+0.26}_{-0.23}$~keV, consistent with the results of \citet{har05} and \citet{hod10}.\\

\noindent
\textbf{4C73.08} is an NLRG with FRII morphology lying in a weak environment with unusual radio features \citep{str13}. We did not detect much emission beyond the host galaxy and could only fit a single-$\beta$ model. \citet{che12} identified 9 group members. Our temperature of 1.4~keV seems a little high for the weak environment, and we found the ICM under-luminous for the temperature compared with the $L_X-T_X$ relation.\\

\noindent
\textbf{DA~240} is a LERG with large FRII lobes. Although it lies in a respectable group of more than 30 galaxies \citep{che11a}, \citet{mil99} failed to detect any X-ray emission from the ICM with a \textit{ROSAT} observation. We likewise detected no emission beyond the host galaxy, and so obtained upper limits on the temperature and luminosity.\\

\noindent
\textbf{NGC~6109} (4C~35.40) is a LERG with a long narrow-angle-tail plume lying in a group of 13 galaxies \citep{mil99}. We found a weak environment beyond the host galaxy, but had insufficient counts to obtain a spectrum. We obtained an estimated temperature of 0.9~keV.\\

\noindent
\textbf{NGC~6251}. This LERG has large FRI plumes extending beyond the imaging chip. It lies in a group of at least 20 galaxies \citep{che11b}, and we detected emission extending beyond the chip. We could not model the background on the outer chips, so obtained the temperature using the single subtraction method with the \textit{Chandra} blank sky files. We obtained a range of temperatures ranging from $\sim1$ to $\sim2.5$~keV from different regions, and used an intermediate result of $1.38^{+0.21}_{-0.03}$~keV, compatible with that found by \citet{eva05} using \textit{XMM-Newton} observations.

We found a much higher luminosity than \citet{eva05} ($0.44\pm{0.07}$ vs $0.07\pm{0.01}\times{10}^{43}$~erg.$s^{-1}$), but a slightly lower luminosity than that which \citet{che11b} estimated from the galaxy velocity dispersions ($\sim0.57\times{10}^{43}$~erg.$s^{-1}$. Our surface brightness profile was wider than that of \citet{eva05} and had a shallower $\beta$. This may be because our \textit{Chandra} observation had a much smaller PSF than the \textit{XMM-Newton} observation used by \citet{eva05}, allowing a more detailed modelling of the surface brightness profile, or because the \textit{Chandra} blank sky files underestimated the background emission. Our result is however close to the $L_X-T_X$ relation.\\

\noindent
\textbf{NGC~7385} (4C~11.71) is a LERG with small FRI plumes. \citet{mil99} found a weak extended environment with a group of 17 galaxies; we found very few counts beyond the host galaxy emission and so obtained upper limits for the temperature and luminosity.

\subsection{2Jy sources}

\textbf{PKS~0034$-$01} (3C~15) is a LERG with small FRII lobes in a weak environment ($B_{gq}\sim90$ -- \citealt{alm13}). Like \citet{rin05}, we found insufficient evidence of ICM emission for analysis so we calculated upper limits for the temperature and luminosity.\\

\noindent
\textbf{PKS~0038+09} (3C~18) is a BLRG with small FRII lobes in a weak environment ($B_{gq}\sim35$ -- \citealt{alm13}). The short observation time gave us insufficient counts for spectral analysis, so we used an estimated temperature of 1.8~keV.\\

\noindent
\textbf{PKS~0043$-$42}. This is a spectroscopic LERG with FRII morphology lying in a weak cluster ($B_{gq}\sim250$ -- \citealt{alm13}) and with evidence of a dusty torus \citep{alm11}. The host galaxy was small in angular extent so we could not model it separately. We found a temperature typical of a group/weak cluster (1.6~keV).\\

\noindent
\textbf{PKS~0213$-$13} (3C~62) is an NLRG with small FRII lobes in a weak environment ($B_{gq}\sim60$ -- \citealt{alm13}). We detected little emission beyond the host galaxy, and had insufficient counts for a double-$\beta$ model or for spectral analysis. Our estimated temperature of 0.85~keV is typical of a weak group.\\

\noindent
\textbf{PKS~0349$-$27} is an NLRG with FRII morphology and extended regions of ionised gas \citep{gri99}, perhaps resulting from a previous merger. We found a temperature typical of a weak group environment (0.86 keV); \citet{alm13} found a stronger $B_{gq}$ than we would expect for that temperature ($B_{gq}\sim200$). \\

\noindent
\textbf{PKS~0404+03} (3C~105) is an NLRG with FRII morphology. \citet{alm13} found that it lay in a weak environment ($B_{gq}\sim80$). The observation time is short, and we found only very slight evidence of ICM emission beyond the host galaxy. We therefore derived upper limits for this source.\\

\noindent
\textbf{PKS~0442$-$28}. This is also an NLRG with FRII morphology. We found no evidence of ICM emission beyond the host galaxy, so derived upper limits for this source. There are however several galaxies close to the host, and \citet{alm13} found a $B_{gq}$ of $\sim450$.\\

\noindent
\textbf{PKS~0620$-$52} is a LERG with FRI plumes lying at a steep angle to each other. It lies in a cluster ($B_{gq}\sim900$ -- \citealt{alm13}). The emission extended beyond the imaging chips, and we were also unable to use a side chip to model the background. We used the \textit{Chandra} blank sky files for the background for both the profile and the spectral analysis. Our temperature of 2.8~keV is typical of a weak cluster.\\

\noindent
\textbf{PKS~0625$-$35}. This source has an optical classification of LERG, but is classified by \citet{wil04} as a possible BL~Lac object and by \citet{gli08} as a LERG. It has FRI morphology and lies in cluster A~3392 ($B_{gq}\sim5000$ -- \citealt{alm13}). We obtained a cluster temperature (3.5~keV). The emission extended beyond the imaging chip so we used the \textit{Chandra} blank sky files for the background for both the profile and the spectral analysis.\\

\noindent
\textbf{PKS~0625$-$53} is a LERG with a FRII morphology lying in a disturbed cluster (A~3391). The emission extended beyond the imaging chip but we were able to use the spectrum from one of the side chips to model the background. We obtained a higher temperature than \citet{fra13} ($6.84^{+0.38}_{-0.34}$ vs $5.21\pm0.03$~keV). This difference is in line with that expected from the reported difference between temperatures obtained using \textit{Chandra} and \textit{XMM-Newton} \citep{sch15}.

This is our hottest ICM, although not the most luminous, and it lies slightly below the $L_X-T_X$ relation -- the \textit{XMM-Newton} temperature of \citet{fra13} would bring the source onto the $L_X-T_X$ relation. \\

\noindent
\textbf{PKS~0806$-$10} (3C~195). The host galaxy of PKS~0806$-$10 may be interacting with a nearby galaxy \citep{ins10}. It is an NLRG with FRII morphology. Although \citet{alm13} found a strong group/weak cluster environment ($B_{gq}\sim600$), we found little emission beyond the host galaxy so used an upper limit.\\

\noindent
\textbf{PKS~0915$-$11} (Hyd~A, 3C~218) is a LERG with FRI morphology lying in cluster Abell~780 ($B_{gq}\sim800$ -- \citealt{alm13}). There is a large shock at about 300~arcsec \citep{sim09}. The emission extended beyond the imaging chips, and we were also unable to use a side chip to model the background so we used the \textit{Chandra} blank sky files for the background for both the profile and the spectral analysis.

We obtained a temperature profile of a very similar shape to that reported by \citet{sim09} using an \textit{XMM-Newton} observation, but with a higher temperature in line with the \textit{Chandra} and \textit{XMM-Newton} difference reported by \citet{sch15}. \citet{har10} obtained a slightly higher temperature for a region nearer the nucleus, but it is compatible with our values for the temperature profile in that region.\\

\noindent
\textbf{PKS~0945+07} (3C~227). There is a dust torus around this BLRG extending to $\sim0.5$~kpc \citep{van10} and an emission line region extending to $\sim100$~kpc \citep{pri97}. It is an FRII lying in a weak environment ($B_{gq}\sim80$ -- \citealt{alm13}). The detected emission does not extend far beyond the host galaxy, so we had insufficient counts for spectral analysis. Our estimated temperature of 1.6~keV is that of a group environment. \\

\noindent
\textbf{PKS~1559+02} (3C~327) is an NLRG with a possible double nucleus and dust lanes \citep{kof00,alm11}. It has FRII radio morphology. \citet{alm13} found a strong group environment ($B_{gq}\sim500$), but we found the ICM emission to be weak with a temperature of only 0.65~keV. The surface brightness profile is very wide and shallow, and we fixed the $\beta$ parameter at the lower limit. Perhaps because of this, we found the luminosity to be high for the temperature compared with that expected from the $L_X-T_X$ relation.\\

\noindent
\textbf{PKS~1648+05} (Her~A, 3C~348) is a LERG with radio characteristics of both FRI and FRII morphologies \citep{sad02}. It has a strong, disturbed environment, a secondary nucleus with a strong shock front \citep{nul05}, and entrained gas and dust filaments that may have come from a now stripped companion galaxy \citep{ode13}. The shock front is clearly visible in the surface brightness profile.

The emission extended beyond the imaging chip but we were able to use the spectrum from the outer regions to model the background. Our temperature of $4.34\pm{0.11}$~keV is that of a moderate cluster, and is consistent with those reported by \citet{nul05} and \citet{giz04} for the unshocked gas (using a shorter \textit{Chandra} observation and a \textit{ROSAT} observation respectively). \citet{har10} obtained a slightly higher temperature from a smaller region nearer the nucleus and crossing the shock front; their temperature was compatible with a similar region in our temperature profile. \citet{alm13} found that PKS~1648+05 is in a strong group ($B_{gq}\sim500$) --- a weaker environment than expected from our temperature.

\citet{giz04} obtained a bolometric luminosity of $4.84\times{10}^{44}$~erg~s$^{-1}$ within a radius of 500~kpc; we obtained a luminosity of $5.9\times{10}^{44}$~erg~s$^{-1}$ within 938~kpc.\\

\noindent
\textbf{PKS~1733$-$56}. This source is a BLRG with FRII morphology lying in a disturbed environment and accreting gas from a merging galaxy \citep{bry02}. We found an unusually wide profile, perhaps a result of the disturbance, with a temperature typical of a group (1.4~keV). \citet{alm13}, however, found a very weak environment ($B_{gq}\sim10$).\\

\noindent
\textbf{PKS~1839$-$48} is a LERG with a double nucleus \citep{alm11}. It has FRI plumes and lies in a rich environment. \citet{alm13} found a $B_{gq}$ of around $\sim1600$, and we obtained a correspondingly high ICM temperature (6~keV).\\

\noindent
\textbf{PKS~1949+02} (3C~403) is an NLRG with X-shaped, FRII morphology lying in a weak environment. We replicated the ISM temperature of 0.24~keV found by \citet{hod10}, but had insufficient counts for spectral analysis of the extended emission. We therefore used an estimated temperature of $0.93^{+0.07}_{-0.12}$~keV. This is a little hotter than the ICM temperature of $0.6\pm{0.2}$ obtained by \citet{hod10}. \citet{alm13} obtained a group-type environment ($B_{gq} \sim120-300$).\\

\noindent
\textbf{PKS~1954$-$55}. This is a LERG with FRI morphology lying in a rich group environment ($B_{gq}\sim500$ -- \citealt{alm13}). The emission extended beyond the imaging chips, but we were able to use a side chip to model the background for the spectral analysis. Our temperature is that of a weak cluster (2.9~keV).\\

\noindent
\textbf{PKS~2211$-$17} (3C~444) is a LERG with cavities around its FRII lobes and a large-scale shock at about 200~kpc \citep{cro11}. It lies in a rich environment ($B_{gq}\sim1300$ --- \citealt{alm13}). We obtained a similar temperature profile to that of \citet{cro11}. For this work, we used a temperature within an annulus excluding the central temperature dip and extending across a wider region, and our temperature and luminosity are consequently a little higher ($4.11^{+0.18}_{-0.18}$ vs $3.5\pm{0.2}$~keV and $2.86^{+0.02}_{-0.03}$ vs $1.0\times{10}^{44}$~erg~s$^{-1}$).\\

\noindent
\textbf{PKS~2221$-$02} (3C~445) is a BLRG with FRII morphology which is thought to have disc winds coming from the central regions \citep{Bra11}. \citet{alm13} found a weak environment ($B_{gq}\sim50$). We had insufficient counts left after subtraction of the \textit{Chandra} blank sky files to obtain a background for the spectral analysis, so we used single background subtraction for the temperature. Our temperature is that of a group (1.1~keV), and is consistent with that found by \citet{hod10}.\\

\noindent
\textbf{PKS~2356$-$61} is an NLRG with FRII lobes in a group environment ($B_{gq}\sim450$ -- \citealt{alm13}). There were insufficient counts left after subtraction of the \textit{Chandra} blank sky files to obtain a background for the spectral analysis. We therefore used single subtraction using a background from the observing chip to obtain the temperature, which was that of a group (1.2~keV). \citet{hod10} found a two-temperature ICM (3.0 and 0.9~keV) but used a much smaller radius for the spectrum. We could not fit a second thermal component with our wider region.

\bsp

\label{lastpage}

\end{document}